\documentclass[aps,prl,twocolumn,titlepage,nofootinbib]{revtex4}
\usepackage{graphicx}
\usepackage{color}
\usepackage{dcolumn}
\usepackage{latexsym}
\newcommand{\angstrom}{\textup{\AA}}
\usepackage[normalem]{ulem}
\usepackage{hyperref,amssymb}
\usepackage{url}
\usepackage{color,soul}
\usepackage{graphicx}
\usepackage{verbatim}
\usepackage{multirow}
\usepackage{amsmath}
\newcommand{\beq}{\begin{eqnarray}}
\newcommand{\eeq}{\end{eqnarray}}
\usepackage{mathrsfs}
\usepackage{float,soul}
\usepackage[dvipsnames]{xcolor}
\usepackage{mathtools}
\usepackage{slashed}
\usepackage{physics}	
\usepackage{graphicx}   
\usepackage{epstopdf}
\usepackage{subfigure}  
\usepackage{hyperref}   
\usepackage{bbold}
\usepackage{wasysym}
\usepackage{feynmp}
\usepackage{hyperref}
\hypersetup{colorlinks,
}

\usepackage[latin1]{inputenc}

\begin{document}

\title{\Large Engineering flat bands in twisted-bilayer graphene\\ away from the magic angle with chiral optical cavities}
\author{Cunyuan Jiang$^{1,2,3}$}\author{Matteo Baggioli$^{1,2,3}$}
\email{b.matteo@sjtu.edu.cn}
\author{Qing-Dong Jiang$^{1,4,5}$}
\email{qingdong.jiang@sjtu.edu.cn}
\address{$^1$ School of Physics and Astronomy, Shanghai Jiao Tong University, Shanghai 200240, China}
\address{$^2$ Wilczek Quantum Center, School of Physics and Astronomy, Shanghai Jiao Tong University, Shanghai 200240, China}
\address{$^3$ Shanghai Research Center for Quantum Sciences, Shanghai 201315,China}
\address{$^4$ Tsung-Dao Lee Institute, Shanghai Jiao Tong University, Shanghai 200240, China}
\address{$^5$ Shanghai Branch, Hefei National Laboratory, Shanghai 201315, China}

\begin{abstract}
Twisted bilayer graphene (TBG) is a recently discovered two-dimensional superlattice structure which exhibits strongly-correlated quantum many-body physics, including strange metallic behavior and unconventional superconductivity. Most of TBG exotic properties are connected to the emergence of a pair of isolated and topological flat electronic bands at the so-called magic angle, $\theta \approx 1.05^{\circ}$, which are nevertheless very fragile. In this work, we show that, by employing chiral optical cavities, the topological flat bands can be stabilized away from the magic angle in an interval of approximately $0.8^{\circ}<\theta<1.3^{\circ}$. As highlighted by a simplified theoretical model, time reversal symmetry breaking (TRSB), induced by the chiral nature of the cavity, plays a fundamental role in flattening the isolated bands and gapping out the rest of the spectrum. Additionally, TRSB suppresses the Berry curvature and induces a topological phase transition, with a gap closing at the $\Gamma$ point, towards a band structure with two isolated flat bands with Chern number equal to $0$. The efficiency of the cavity is discussed as a function of the twisting angle, the light-matter coupling and the optical cavity characteristic frequency. Our results demonstrate the possibility of engineering flat bands in TBG using optical devices, extending the onset of strongly-correlated topological electronic phases in moir\'e superlattices to a wider range in the twisting angle.
\end{abstract}
\maketitle
\color{blue}\textit{Introduction}\color{black} -- Controlling and engineering quantum phases of matter is a central task in condensed matter physics. Inspired by the original discovery of single-layer graphene \cite{doi:10.1126/science.1102896}, two-dimensional (2D) materials have emerged as a versatile platform to realize strongly-correlated physics in quantum many body systems \cite{2dreviewneto}. Recently, unconventional superconductivity was discovered in twisted bilayer graphene (TBG), a two dimensional superlattice where one layer of graphene is stacked on top of another at a special magic twisting angle, \textit{i.e.}, $\theta \approx 1.05^{\circ}$ \cite{yuancao2018,Yankowitz_TBG_2019,Lu_TBG_2019,Saito_TBG_2020}. Galvanized by this breakthrough, several other stacked two-dimensional systems that host exotic superconductivity, such as twisted multilayer graphene, have been revealed \cite{Chen_ABC1_2019,Chen_ABC2_2019,Cao_TTG_2021,Tsai_TTG_2021,Cao_mutiLG_2022}. While the underneath physical mechanism of superconductivity in twisted 2D systems is still under debate \cite{PhysRevX.8.031089,PhysRevX.8.041041,PhysRevLett.121.257001,PhysRevLett.121.087001,PhysRevLett.124.167002,PhysRevLett.129.047601,PhysRevLett.121.217001,PhysRevB.99.121407}, it is clear that the isolated electronic flat bands appearing at the magical angle play an essential role. Besides superconductivity, flat bands are also indispensable for the emergence of strongly-correlated insulating states and the strange-metal phase near the superconducting dome in the phase diagram of TBG, which closely mimics that of cuprate superconductors \cite{cao2018correlated, Cao_TBG_strange_2020,liu2021tuning,PhysRevLett.124.097601,PhysRevX.8.031087,PhysRevB.98.075154,PhysRevB.98.045103,PhysRevB.98.075109,pizarro2019nature}.

However, despite being a promising platform for studying strongly correlated physics, the unavoidable and uncontrollable non-uniformity of the twist angle across the sample, and the consequent difficulty in keeping the twist angle at its magic value, prevented a wide realization of these phenomena \cite{parkcao2022,PhysRevResearch.2.023325}. More precisely, since the magical-angle configuration is unstable, a little offset (around $\pm 0.1^{\circ}$) of the twisting angle easily destroys most of the emergent exotic properties of TBG. In this regard, one of the most important challenges in the field is therefore to achieve superconductivity at non-magic values of the twisting angle. To achieve this final goal, it is desirable to realize a primary step, namely to create and stabilize electronic flat bands in a wider range of the twisting angle \cite{PhysRevB.100.035448,choi2019intrinsic,liu2020tunable,PhysRevLett.120.266402,PhysRevLett.123.096802}. This will be the main purpose of our work. 

\begin{figure}
    \centering
    \includegraphics[width=0.9\linewidth]{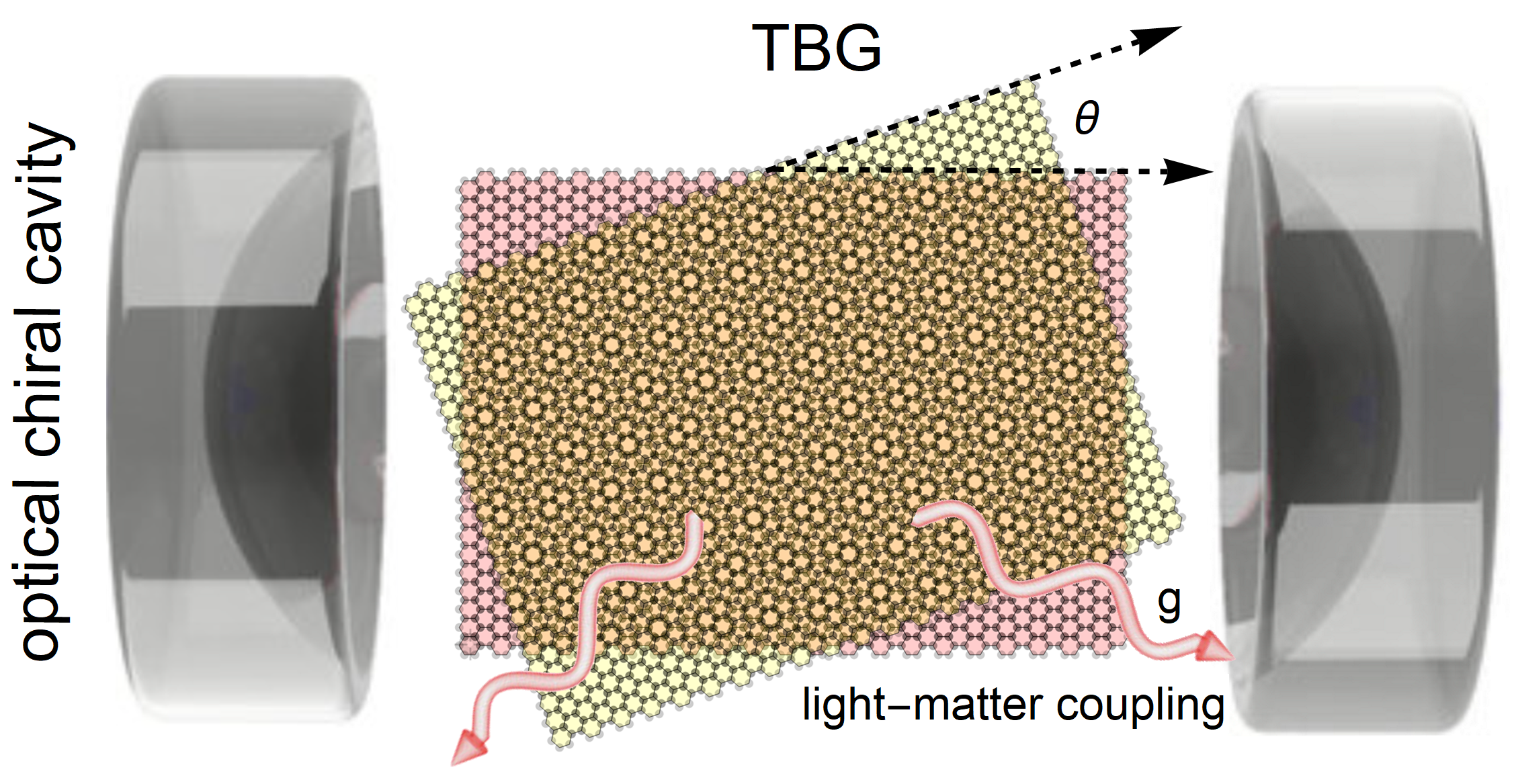}

    \vspace{0.1cm}
    
    \includegraphics[width=0.7\linewidth]{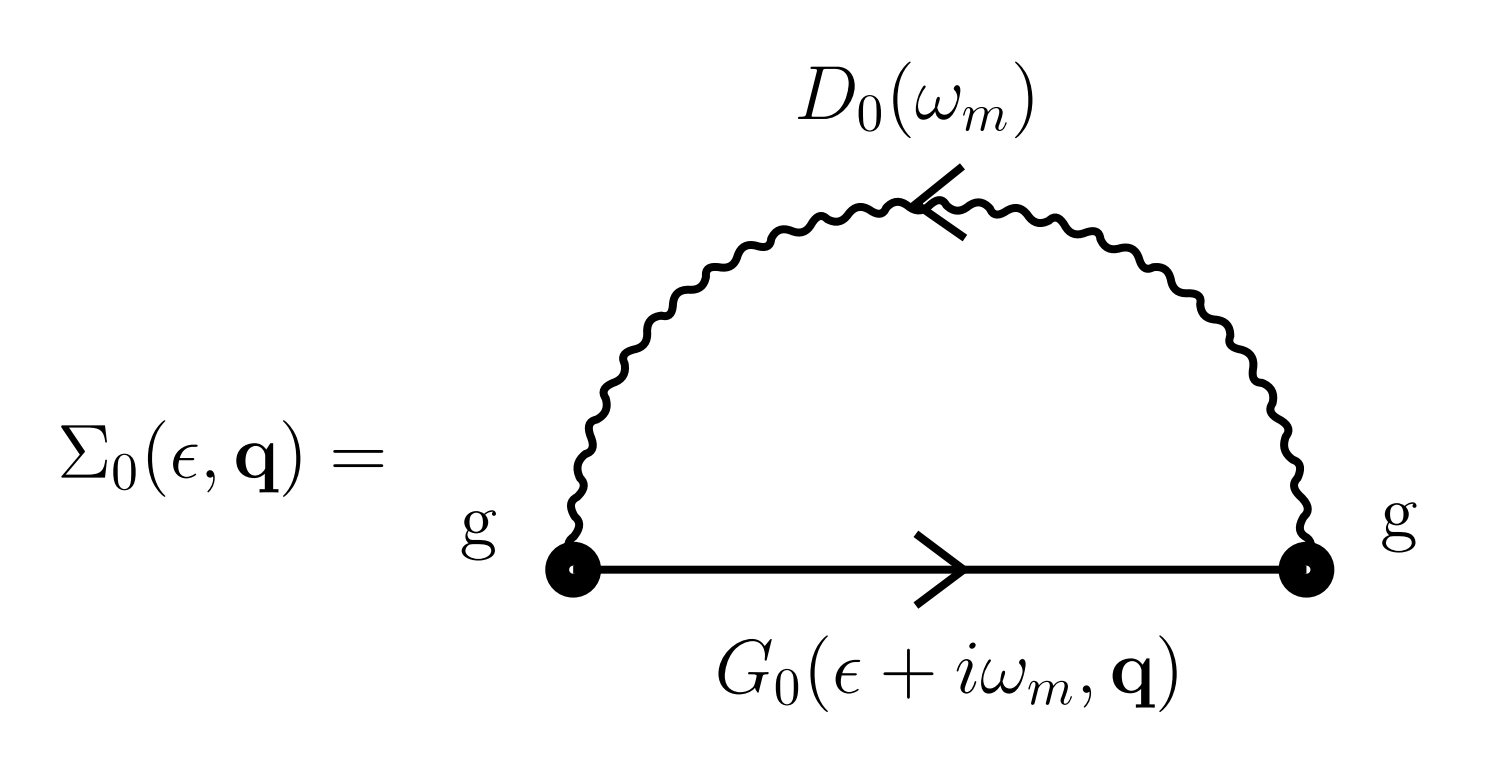}
    \caption{\textbf{Top panel.} A cartoon of the optical setup considered in this work. Two skewed sheets of graphene are stacked on top of each other with a twisting angle $\theta$ creating a characteristic moir\'e pattern. They are then put inside a chiral optical cavity with a light-matter coupling $g$ and a characteristic frequency $\omega_c$. \textbf{Bottom panel.} The dominant Feynman diagram describing light-matter interactions in the optical chiral cavity and giving rise to the self-energy $\Sigma_0(\epsilon,\mathbf{q})$.}
    \label{fig:0}
\end{figure}

In this Letter, we propose a new method to engineer stable flat bands at non-magic angles by embedding twisted-bilayer graphene in a vacuum chiral cavity (see top panel in Figure \ref{fig:0} for a cartoon of the setup). 
Using vacuum cavities to control materials and molecules has emerged as a fruitful playground connecting quantum optics to condensed matter and chemistry \cite{hubener2021engineering,schlawin2022cavity,bloch2022strongly,PhysRevApplied.18.044011,schafer2018ab,bacciconi2023first,mercurio2023photon}. Floquet methods have also been proposed to engineer electronic properties in TBG \cite{PhysRevB.101.241408,PhysRevResearch.1.023031} and have been experimentally applied in different contexts, \textit{e.g.} \cite{doi:10.1126/science.1239834}. However, because of the external electromagnetic radiation which drives the system out of equilibrium, this second route inevitably heats up the system, destroying quantum coherence and inducing transient phenomena away from thermal equilibrium states. Therefore, it is unclear whether a stabilization of the flat bands in TBG using Floquet methods would preserve the related superconductivity properties. In this sense, at least in the case of TBG, optical methods might be superior. In the past, the usage of vacuum cavities has been proposed to design material conductivity \cite{rokaj2022free,moddel2021casimir}, unconventional superconductivity \cite{sentef2018cavity,schlawin2019cavity,curtis2019cavity,thomas2019exploring}, topological properties \cite{PhysRevB.104.155307,espinosa2014semiconductor,PhysRevB.99.235156,appugliese2022breakdown}, and even chemical reactivity \cite{galego2019cavity,galego2015cavity}. Some of these proposals have been already successfully realized experimentally.

A fundamental property of vacuum chiral cavities is that time-reversal symmetry is broken without the need of an external driving \cite{hubener2021engineering,schlawin2022cavity}. The same effect can be achieved using hBN encapsulation (\textit{e.g.}, \cite{Long2022}). Time-reversal symmetry breaking is essential since, as we will see, quantum fluctuations alone can not significantly influence the electronic bands in TBG. In single-layer graphene, a band gap can be induced by quantum fluctuations in a chiral cavity as well \cite{PhysRevB.99.235156}. However, the effect is too small to be directly observed due to the large bandwidth. As we will demonstrate, the situation is different in TBG near the magic angle, where the small bandwidth enables time-reversal symmetry-broken quantum fluctuations to play a significant role. 

In recent years, a number of works have realized the vital impact of symmetry breaking on quantum-fluctuations-related phenomena, such as anomalous Casimir effects \cite{butcher2012casimir,jiang2019axial}, topological gap generation \cite{espinosa2014semiconductor,PhysRevB.99.235156}, angular-momentum-dependent spectral shift \cite{jiang2023angular}, and selection of chiral molecules in chemical reactions \cite{PhysRevLett.131.223601,riso2023strong,vu2023enhanced}. A recent work by one of us \cite{jiang2019quantum} highlighted the combined power of symmetry breaking and quantum fluctuations, proving that symmetry breaking effects can be transmitted from a material to its vicinity by vacuum quantum fluctuations. In this scenario, the vacuum in proximity of a material with broken symmetries is referred to as its ``quantum atmosphere".

In this Letter, we investigate the band renormalization of TBG due to the time-reversal symmetry broken quantum fluctuations in a chiral cavity. We start from a faithful tight-binding model of TBG and calculate the one-loop self-energy induced by the light-matter coupling. The bottom panel of Figure \ref{fig:0} displays the specific Feynman diagram considered. We find that, for experimentally realizable values of the light-matter coupling and cavity frequency, the topological flat bands in TBG can be stabilized away from the magic angle in an interval of approximately $0.8^{\circ}<\theta<1.3^{\circ}$. Our derivation and calculations can be directly generalized to other twisted 2D systems.

\color{blue}\textit{Setup and methods}\color{black} -- To set the stage, we model the Hamiltonian of the combined system, TBG and cavity, as follows:
\begin{eqnarray}
\hat H=\hat H_{\text{TBG}}(\mathbf{q}-e\boldsymbol{\hat A})+\hbar \omega_c \hat a^\dagger \hat a,
\end{eqnarray}
where $H_{\text{TBG}}(\mathbf{q})$ represents the TBG Hamiltonian in reciprocal space, and $\omega_c$ is the cavity photonic mode frequency. TBG and cavity photonic modes are coupled through Peierls substitution $\mathbf q\mapsto \mathbf{q}-e\boldsymbol{\hat A}$, where $\boldsymbol{\hat A}$ can be expressed in terms of photonic creation and annihilation operators, \textit{i.e.}, $\hat{\bold A}=A_0\left(\boldsymbol \varepsilon^* \hat a^\dagger+\boldsymbol\varepsilon \hat a\right)$. Here, $\boldsymbol \varepsilon$ is the polarization tensor of the cavity photonic modes and $A_0=\sqrt{\frac{\hbar}{2\epsilon_0 V\omega_c}}$ is the mode amplitude in terms of the cavity volume $V$. We focus on chiral cavities where the photonic polarization is given by $\boldsymbol\varepsilon=\frac{1}{\sqrt{2}}\left(\mathbf{e_x}+i\mathbf{e_y}\right) $, with $\mathbf{e_{x(y)}}$ the unit vector in the x(y)-direction. Our setup can be straightforwardly generalized to the multi-mode case.

To be concrete, let us consider the effective tight-binding Hamiltonian $H_{\text{TBG}}(\mathbf{q})$ \cite{handbook}:
\begin{equation}
    \left(\begin{matrix}
    H_1 (\mathbf{q}) & T_{\mathbf{q}_b} & T_{\mathbf{q}_{tr}} & T_{\mathbf{q}_{tl}} & \cdots\\ T_{\mathbf{q}_b}^\dagger &  H_2 (\mathbf{q}-\mathbf{q}_b) & 0 & 0 & \cdots\\ T_{\mathbf{q}_{tr}}^\dagger & 0 & H_2 (\mathbf{q}-\mathbf{q}_{tr}) & 0  & \cdots\\ T_{\mathbf{q}_{tl}}^\dagger & 0 & 0 & H_2 (\mathbf{q}-\mathbf{q}_{tl}) & \cdots \\ \vdots & \vdots & \vdots & \vdots & \ddots
    \end{matrix}\right)\label{hmain}
\end{equation}
where $\mathbf{q}$ is the wave-vector, and $H_{1,2}(\mathbf{q})$ indicate the Hamiltonian of the top/bottom layer respectively. Moreover, we have defined:
\begin{align}
   & \mathbf{q}_b=\dfrac{1}{3}(\mathbf{b}_1^m-\mathbf{b}_2^m),\quad \mathbf{q}_{tr}=\dfrac{1}{3}(\mathbf{b}_1^m+2\mathbf{b}_2^m),\nonumber\\
   &\mathbf{q}_{tl}=\dfrac{1}{3}(-2\mathbf{b}_1^m-\mathbf{b}_2^m),
\end{align}
where $\mathbf{b}_1^m$ and $\mathbf{b}_2^m$ are moir\'e reciprocal vectors. The twisting angle $\theta$ is hidden in these vectors; see Fig.6.11 in Ref.\cite{handbook} for more details. The hopping matrix elements are given by,
\begin{align}
   & T_{\mathbf{q}_{b}}= t\left(\begin{matrix}
    u & u' \\ u' & u
    \end{matrix}\right),\quad
    T_{\mathbf{q}_{tr}}= t\left(\begin{matrix}
    u e^{i \phi} & u' \\ u' e^{-i \phi} & u e^{i \phi}
    \end{matrix}\right),\nonumber\\ 
   & \qquad \qquad T_{\mathbf{q}_{tl}}= t\left(\begin{matrix}
    u e^{-i \phi} & u' \\ u' e^{i \phi} & u e^{-i \phi}
    \end{matrix}\right),
\end{align}
where the various parameters have been fixed to $u=0.817$, $u'=1$, $\phi=2\pi/3$ and $t=0.11$ is the hopping parameter. This choice takes into account the different interlayer coupling strength of AA and AB stacked regions \cite{PhysRevX.8.031087} due to the surface relaxation effects and the consequent atomic corrugation \cite{PhysRevB.90.155451,PhysRevB.99.195419}. For more details about the TBG Hamiltonian we refer to the Supplementary Information (SI).
Once the effective Hamiltonian is known, the bare electron propagator can be obtained using \cite{PhysRevB.99.235156}
\begin{equation}
    G_0^{-1}(\epsilon,\textbf{q})=\left[\left(\epsilon+i 0^+\right)\mathbf{I}-H_{\text{TBG}}(\textbf{q})\right]^{-1}.
\end{equation}
The full electron propagator, taking into account the interactions with the vacuum cavity, can then be derived from the Dyson equation
\begin{equation}
    G^{-1} (\epsilon,\mathbf{q})=G_0^{-1} (\epsilon,\mathbf{q})-\Sigma_0 (\epsilon,\mathbf{q}),\label{propagator}
\end{equation}
where $\Sigma_0$ is the self-energy (see bottom panel of Fig.\ref{fig:0}) given by
\begin{equation}
    \Sigma_0 (\epsilon,\mathbf{q}) =-\dfrac{g^2}{\beta}\sum_{m=1}^\infty G_0 (\epsilon+i\omega_m,\mathbf{q}) D_0 (\omega_m).\label{sum}
\end{equation}
Here, $\omega_m=2 \pi m k_B T$ is the $m$th Matsubara frequency and $g=v_F e A_0$ denotes the electron-photon coupling strength with $v_F$ the Fermi velocity of monolayer graphene and $e$ the electromagnetic coupling. For convenience, we define the dimensionless coupling $\tilde g\equiv g/k_B T$. Finally, $D_0 (\omega_m)$ is the photon propagator given by
\begin{equation}
    D_0 (\omega_m)= \left(\begin{matrix}
    \dfrac{-1}{i\omega_m +\omega_c} & 0 & \cdots \\ 0 & \dfrac{1}{i\omega_m -\omega_c} & \cdots \\ \vdots & \vdots & \ddots
    \end{matrix}\right),
\end{equation}
with $\omega_c$ the cavity frequency. It should be noticed here that all quantities $G_0$, $G$, $D_0$, $\Sigma_0$ are matrices with the same dimension of the effective TBG Hamiltonian. With the full propagator in Eq.\eqref{propagator}, the spectral function, giving the renormalized electronic band structure, can be calculated from
\begin{equation}
    A(\epsilon,\textbf{q})=-\frac{1}{\pi}\mathrm{Im}\mathrm{Tr}G(\epsilon,\textbf{q}).
\end{equation}
More details about the TBG Hamiltonian, the structure in reciprocal space and the numerical methods employed can be found in the SI.\\ 
\begin{figure}
    \centering
    \includegraphics[width=\linewidth]{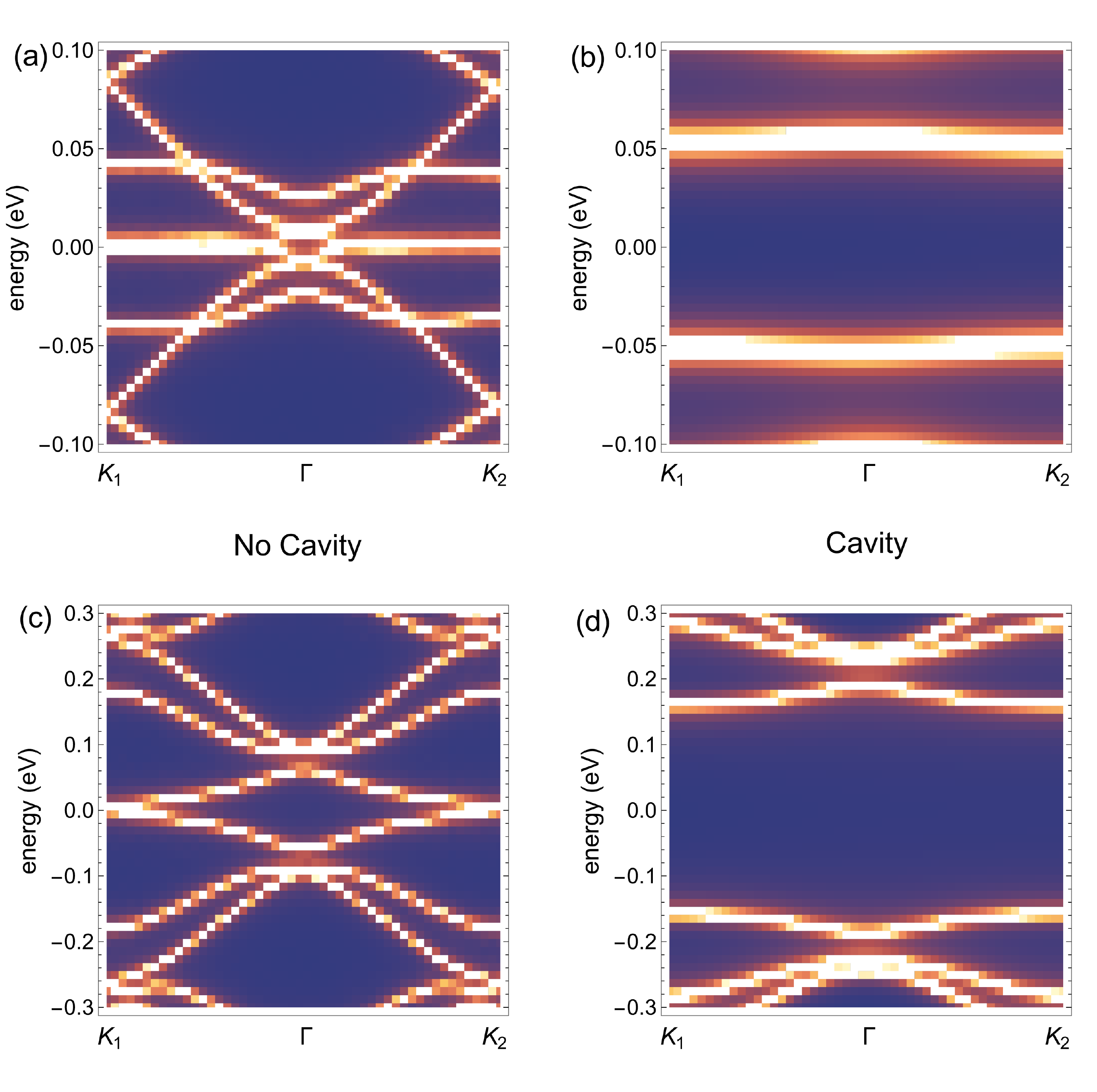}
    \caption{The spectral function $A(\epsilon,q)$ showing the band structure of TBG with different light-matter coupling strength $g$ and twisting angle $\theta$. \textbf{(a)}: $\theta=0.8^{\circ},\,\tilde g=0$; \textbf{(b)}: $\theta=0.8^{\circ},\, \tilde g=2$; \textbf{(c)}: $\theta=1.5^{\circ},\,\tilde g=0$; \textbf{(d)}: $\theta=1.5^{\circ},\, \tilde g=2.5$. The optical cavity characteristic frequency is fixed to $\omega_c=0.3$eV.}
    \label{fig:1}
\end{figure}

\color{blue}\textit{Electronic spectrum}\color{black} -- It is well known that TBG exhibits a pair of topological flat bands at the magic angle, $\theta \approx 1.05^{\circ}$ \cite{doi:10.1073/pnas.1108174108,PhysRevB.82.121407,Andrei2020}, which play a key role for the underlying strongly-correlated physics. In panels (a) and (c) of Fig.\ref{fig:1}, we show two examples of the electronic spectrum of TBG at respectively $\theta=0.8^{\circ}$ and $\theta=1.5^{\circ}$. As already mentioned, no isolated flat bands are present anymore in the electronic spectrum just moving of $\pm 0.3^{\circ}$ from the magic angle. In other words, the flat bands are very fragile and sensitive to the twisting angle. In panels (b) and (d) of Fig.\ref{fig:1}, we show the same electronic spectra in presence of the chiral optical cavity, with a coupling $\tilde g=2$ and a characteristic frequency $\omega_c=0.3$ eV. The strength of this coupling corresponds to a micrometer-sized cavity, well within experimental reach, as demonstrated by a similar experimental value in \cite{Song2005}. A pair of nearly-flat bands re-appear away from the magic angle thanks to the coupling to the chiral cavity. Importantly, the two bands are not anymore degenerate as their energy is shifted from the Fermi energy and grows with the light-matter coupling $\tilde g$. This is a direct consequence of the breaking of time reversal symmetry induced by the chiral cavity. At the same time, the other bands, similarly to the case of the Dirac cone in monolayer graphene (see SI), are also gapped away as a result of the same symmetry breaking pattern.

\begin{figure}
    \centering
    \includegraphics[width=\linewidth]{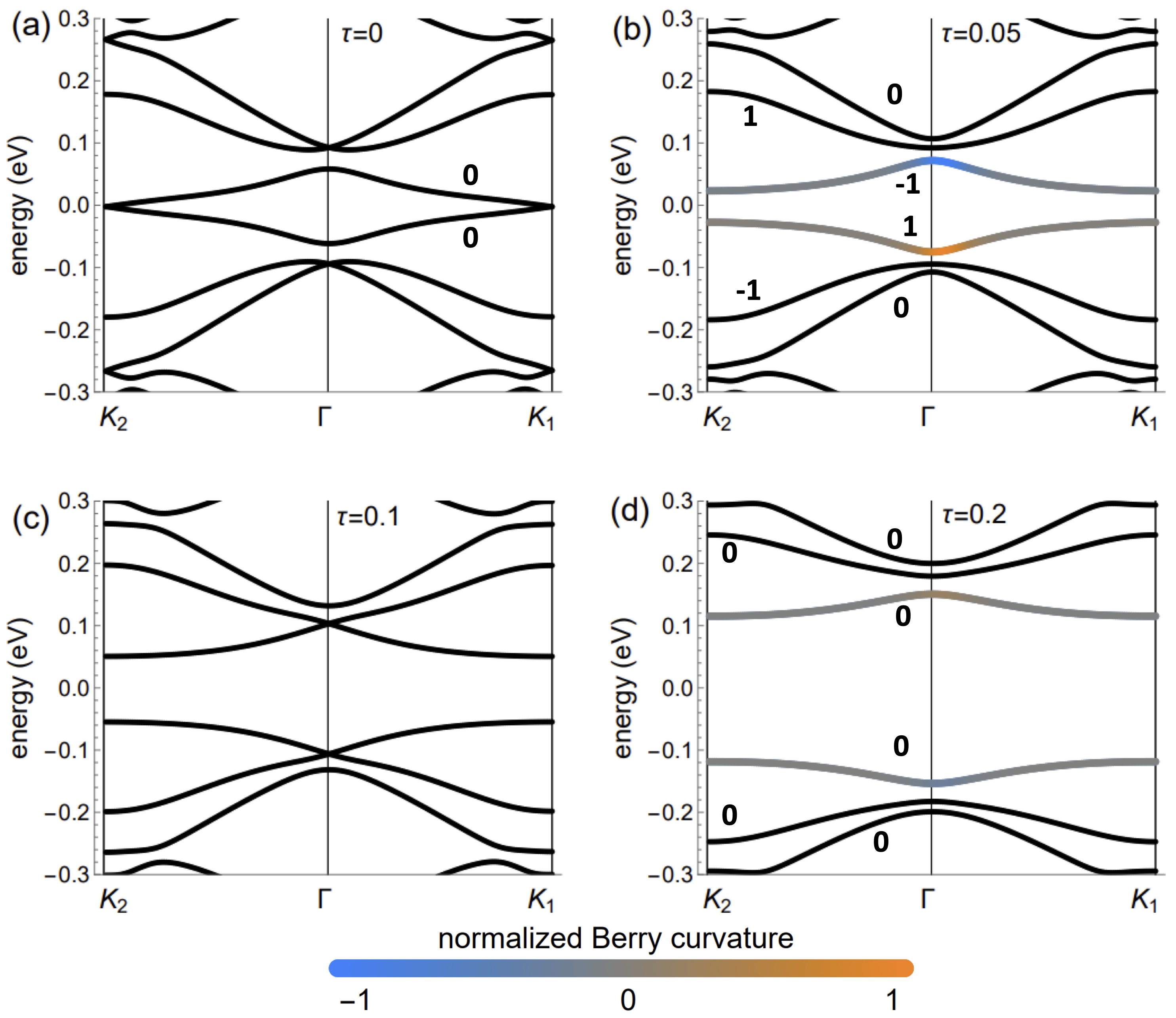}
    \caption{The band structure of TBG at $\theta=1.5^{\circ}$ obtained from the deformed Hamiltonian in Eq.\eqref{h+t}. Panels \textbf{(a)}, \textbf{(b)}, \textbf{(c)}, \textbf{(d)} respectively correspond to an increasing value for the time reversal breaking parameter $\tau=0,0.05,0.15,0.2$. The color scheme on top of the electronic bands in panels indicates the value of the reduced Berry curvature, and the integer numbers the Chern number of the lowest bands.}
    \label{fig:addpaulimatrix}
\end{figure}

\color{blue}\textit{Theoretical model \& topological properties}\color{black} -- The emergence of the two isolated quasi-flat bands and their energy splitting, shown in Fig.\ref{fig:1}, are intimately connected to the chiral nature of the optical cavity, which plays a fundamental role in this regard. To explicitly prove this statement, we construct a simplified analytical model which, as we will see, possesses all the minimal ingredients to describe our setup. In order to model the effects of the chiral cavity on TBG, we consider the following deformed Hamiltonian:
\begin{equation}
    H_{\text{TBG}+\tau}(\mathbf{q})=H_{\text{TBG}}(\mathbf{q})
    +\tau\left(\begin{matrix}
    \sigma_z & 0 & 0 & 0 & \cdots\\ 0 &  \sigma_z & 0 & 0 & \cdots\\ 0 & 0 & \sigma_z & 0  & \cdots\\ 0 & 0 & 0 & \sigma_z & \cdots \\ \vdots & \vdots & \vdots & \vdots & \ddots
    \end{matrix}\right),
    \label{h+t}
\end{equation}
where the coupling $\tau$ parameterizes the breaking of time-reversal symmetry on top of the original TBG Hamiltonian in Eq.\eqref{hmain}. This is unlikely the most general deformation which breaks time-reversal symmetry, but it will be sufficient to qualitatively reproduce the numerical results displayed in Fig.\ref{fig:1} and identify the main underlying physical principle behind them.
By diagonalizing the above Hamiltonian $H_{\text{TBG}+\tau}(\mathbf{q})$, the band structure with broken time reversal symmetry can be obtained. The results are shown in Fig.\ref{fig:addpaulimatrix} for different strength of the time-reversal symmetry breaking $\tau$. As clearly demonstrated, the effects of $\tau$ is to gap away the higher energy bands and create a pair of isolated quasi-flat bands wth non-degenerate energy. At least at a qualitative level, these results are in perfect agreement with the more realistic scenario of TBG in a chiral cavity shown in Fig.\ref{fig:1}, where the light-matter coupling $\tilde g$ plays an analogous role of the phenomenological parameter $\tau$ in Eq.\eqref{h+t}. This simplified but tractable analytical model highlights the fundamental role of time reversal symmetry breaking, induced by the chiral cavity, in stabilizing the flat band of TBG. In order to characterize better the effects of the cavity, and of time-reversal breaking, on the electronic bands in TBG, we have computed the Berry curvature and the corresponding topological Chern number, that are geometrical properties of an energy band connected to how eigenstates evolve as a local function of parameters \cite{RevModPhys.82.1959} (see SI for details).

At $\tau=0$, all the bands are topologically trivial because of symmetry constraints. By introducing time-reversal symmetry breaking, a gap at the $K$ points opens between the two lowest bands and at the $\Gamma$ point between the two consecutive ones. Qualitatively, for the two lowest bands, this mechanism is very similar to what is observed in single-layer graphene \cite{PhysRevResearch.1.023031}. The bands becomes topological, with a finite $\pm 1$ Chern number and a non-trivial Berry curvature exhibiting dipolar structure (Berry curvature dipole) \cite{PhysRevLett.115.216806}. By increasing $\tau$ further, the gap at the $\Gamma$ point closes and the system undergoes a topological phase transition towards a topologically trivial state in which all electronic bands display a vanishing Chern number, as for $\tau=0$. Despite the state is topologically trivial, the Berry curvature does not vanish identically in the whole Brillouin zone but displays an interesting dipolar structure. To the best of our knowledge, this topological phase transition in TBG induced by time reversal symmetry breaking was not discussed before. Importantly, the same behavior is also directly observed when TBG is coupled to the optical cavity (see SI), and it potentially represents a universal feature of TBG with broken time reversal symmetry that deserves further investigation.

\color{blue}\textit{Phase diagram and quasi-flat bands}\color{black} -- In order to explore the effects of the chiral cavity in more detail, we have performed an extensive analysis of the band structure for different values of $\theta$ and $\tilde g$ covering a wide range of the phase diagram around the magic-angle value. To give a quantitative estimate of the flatness of the bands, we define the energy bandwidth parameter
\begin{equation}
    \Delta \epsilon(g,\theta)\equiv \epsilon_{b}^{\text{max}}(g,\theta)-\epsilon_{b}^{\text{min}}(g,\theta)
\end{equation}
which quantifies the variation of the energy along the isolated nearly-flat bands. As a reference, a completely flat band would correspond to $\Delta \epsilon(g,\theta)=0$. 
\begin{figure}
    \centering
    \includegraphics[width=0.9\linewidth]{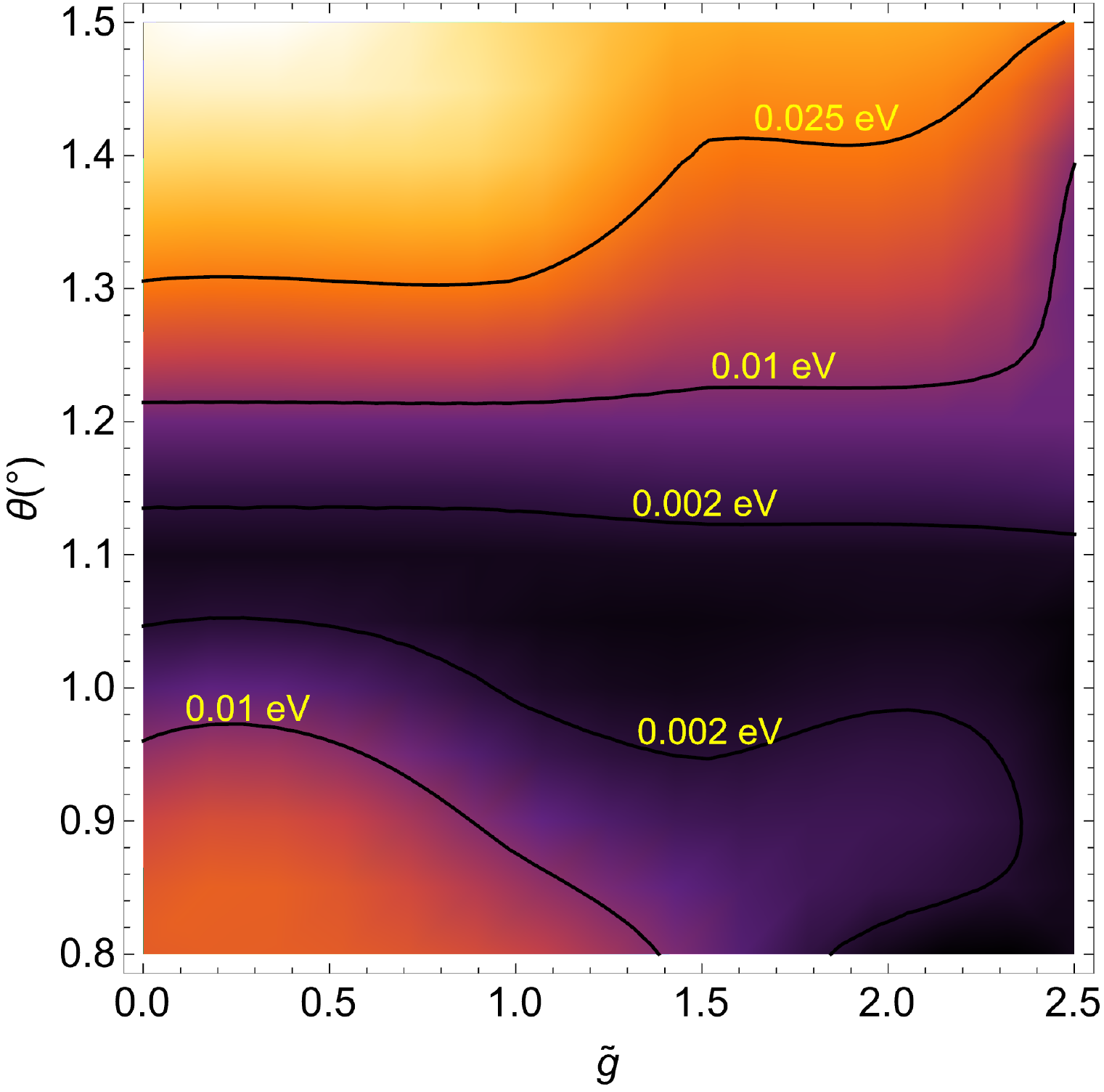}
    \caption{The value of the energy bandwidth $\Delta \epsilon(\tilde g,\theta)$ for different $\tilde g$ and $\theta$. Darker color corresponds to a flatter band. The black solid lines indicates a few constant $\Delta \epsilon$ values as reference. The optical cavity frequency is set to $\omega_c=0.3$eV.}
    \label{fig:2}
\end{figure}
Our results are shown in Fig.\ref{fig:2} in the interval $0.8^{\circ}<\theta<1.5^{\circ}$ and $0<\tilde g<2.5$ for a fixed and reasonable value of the cavity characteristic frequency $\omega_c=0.3$ eV. By increasing the value of the dimensionless coupling $\tilde g$, the energy bandwidth becomes smaller and therefore the isolated bands can be flattened away from the magic angle. As expected from simple intuition, the isolated bands can be flattened more efficiently for angles which are closer to the magic value. Interestingly, we find that it is easier to flatten the bands for angles smaller than the magic one compared to angles larger than the latter. 

In general, we observe that ``quasi-flat'' bands, with a variation of the energy within $0.01$eV, can be easily tuned using reasonable values of the light-matter coupling $\tilde{g}\sim \mathcal{O}(1)$ for angles of $\pm 0.3^{\circ}$ away from the magic value $\approx 1.05^{\circ}$. Notice that, using the definition $g=\sqrt{\frac{\hbar v_F^2 e^2}{2 \epsilon_0 V \omega_c}}$ and considering an optical cavity of volume $V=(1~{\rm \mu m})^3$, frequency of the order $\omega_c\approx 10^{-1} ~{\rm eV}$, and at a temperature of $\approx 1$K, this corresponds to a dimensionful light-matter coupling $g$ of the order of $\approx 10^{-4} {\rm eV}$, which is certainly within experimental reach, as corroborated by existing literature \cite{Song2005,PhysRevB.90.205309,FriskKockum2019}. Further elaboration on the experimental feasibility is provided in the SI, which
includes Refs. \cite{scalari2013ultrastrong,PhysRevA.107.L021501,jarc2023cavity}.

\color{blue}\textit{Conclusions}\color{black} -- In this Letter, we have revealed the possibility of extending the onset of topological flat-bands in twisted bilayer graphene away from the magic angle by using optical chiral cavities. We have demonstrated that the effects of light-matter coupling can stabilize and flatten the topological flat bands for a large range of the twisting angle without the need of fine-tuning. Using physical values for the optical cavity frequency and the strength of the light-matter coupling, we have estimated that quasi-flat bands can be achieved at least in an interval of $0.8^{\circ}<\theta<1.3^{\circ}$. From a theoretical point of view, taking advantage of a simplified analytical model, we have identified the breaking of time-reversal symmetry as the fundamental ingredient behind the achieved flattening, and the responsible of a previously overlooked topological phase transition.

One immediate future task is to verify whether all the interesting strongly-correlated physics related to the topological flat bands survive in presence of the chiral cavity or how that is modified. For example, it would be interesting to understand further the effects of time-reversal symmetry breaking on the emergent exotic superconductivity of TBG. Additionally, several works have emphasized the importance of quantum geometry in determining light-matter coupling strength in TBG. Therefore, a second task is to explore the role of quantum geometry in cavity-induced band renormalization in TBG \cite{PhysRevResearch.1.023031,PhysRevB.104.064306,tai2023quantum}.
The last and more pressing point is to verify our theoretical predictions within an experimental setup. Following our preliminary estimates (see more details in the SI), we conclude that the results shown in this Letter might already be within experimental reach.

In general, we expect the combination of twistronics and photonics engineering to become a powerful platform to study strongly-correlated electronic systems and topological matter beyond the case of twisted bilayer graphene. 

\color{blue}{\it Acknowledgments} \color{black} --  We would like to thank Dario Rosa for collaboration at an early stage of this project. We appreciate helpful discussions with Jinhua Gao and Miao Liang. We would like to thank the anonymous Referees for several constructive comments. CJ and MB acknowledge the support of the Shanghai Municipal Science and Technology Major Project (Grant No.2019SHZDZX01). MB acknowledges the support of the sponsorship from the Yangyang Development Fund. Q.-D. Jiang was sponsored by National Natural Science Foundation of China (NSFC) under Grant No. 23Z031504628, Pujiang Talent Program 21PJ1405400, Jiaoda 2030 program WH510363001-1, and the Innovation Program for Quantum Science and Technology Grant No.2021ZD0301900.

\bibliographystyle{apsrev4-1}
\bibliography{ref}

\begin{thebibliography}{81}%
\makeatletter
\providecommand \@ifxundefined [1]{%
 \@ifx{#1\undefined}
}%
\providecommand \@ifnum [1]{%
 \ifnum #1\expandafter \@firstoftwo
 \else \expandafter \@secondoftwo
 \fi
}%
\providecommand \@ifx [1]{%
 \ifx #1\expandafter \@firstoftwo
 \else \expandafter \@secondoftwo
 \fi
}%
\providecommand \natexlab [1]{#1}%
\providecommand \enquote  [1]{``#1''}%
\providecommand \bibnamefont  [1]{#1}%
\providecommand \bibfnamefont [1]{#1}%
\providecommand \citenamefont [1]{#1}%
\providecommand \href@noop [0]{\@secondoftwo}%
\providecommand \href [0]{\begingroup \@sanitize@url \@href}%
\providecommand \@href[1]{\@@startlink{#1}\@@href}%
\providecommand \@@href[1]{\endgroup#1\@@endlink}%
\providecommand \@sanitize@url [0]{\catcode `\\12\catcode `\$12\catcode
  `\&12\catcode `\#12\catcode `\^12\catcode `\_12\catcode `\%12\relax}%
\providecommand \@@startlink[1]{}%
\providecommand \@@endlink[0]{}%
\providecommand \url  [0]{\begingroup\@sanitize@url \@url }%
\providecommand \@url [1]{\endgroup\@href {#1}{\urlprefix }}%
\providecommand \urlprefix  [0]{URL }%
\providecommand \Eprint [0]{\href }%
\providecommand \doibase [0]{http://dx.doi.org/}%
\providecommand \selectlanguage [0]{\@gobble}%
\providecommand \bibinfo  [0]{\@secondoftwo}%
\providecommand \bibfield  [0]{\@secondoftwo}%
\providecommand \translation [1]{[#1]}%
\providecommand \BibitemOpen [0]{}%
\providecommand \bibitemStop [0]{}%
\providecommand \bibitemNoStop [0]{.\EOS\space}%
\providecommand \EOS [0]{\spacefactor3000\relax}%
\providecommand \BibitemShut  [1]{\csname bibitem#1\endcsname}%
\let\auto@bib@innerbib\@empty
\bibitem [{\citenamefont {Novoselov}\ \emph {et~al.}(2004)\citenamefont
  {Novoselov}, \citenamefont {Geim}, \citenamefont {Morozov}, \citenamefont
  {Jiang}, \citenamefont {Zhang}, \citenamefont {Dubonos}, \citenamefont
  {Grigorieva},\ and\ \citenamefont {Firsov}}]{doi:10.1126/science.1102896}%
  \BibitemOpen
  \bibfield  {author} {\bibinfo {author} {\bibfnamefont {K.~S.}\ \bibnamefont
  {Novoselov}}, \bibinfo {author} {\bibfnamefont {A.~K.}\ \bibnamefont {Geim}},
  \bibinfo {author} {\bibfnamefont {S.~V.}\ \bibnamefont {Morozov}}, \bibinfo
  {author} {\bibfnamefont {D.}~\bibnamefont {Jiang}}, \bibinfo {author}
  {\bibfnamefont {Y.}~\bibnamefont {Zhang}}, \bibinfo {author} {\bibfnamefont
  {S.~V.}\ \bibnamefont {Dubonos}}, \bibinfo {author} {\bibfnamefont {I.~V.}\
  \bibnamefont {Grigorieva}}, \ and\ \bibinfo {author} {\bibfnamefont {A.~A.}\
  \bibnamefont {Firsov}},\ }\href {\doibase 10.1126/science.1102896} {\bibfield
   {journal} {\bibinfo  {journal} {Science}\ }\textbf {\bibinfo {volume}
  {306}},\ \bibinfo {pages} {666} (\bibinfo {year} {2004})}\BibitemShut
  {NoStop}%
\bibitem [{\citenamefont {Novoselov}\ \emph {et~al.}(2016)\citenamefont
  {Novoselov}, \citenamefont {Mishchenko}, \citenamefont {Carvalho},\ and\
  \citenamefont {Neto}}]{2dreviewneto}%
  \BibitemOpen
  \bibfield  {author} {\bibinfo {author} {\bibfnamefont {K.~S.}\ \bibnamefont
  {Novoselov}}, \bibinfo {author} {\bibfnamefont {A.}~\bibnamefont
  {Mishchenko}}, \bibinfo {author} {\bibfnamefont {A.}~\bibnamefont
  {Carvalho}}, \ and\ \bibinfo {author} {\bibfnamefont {A.~H.~C.}\ \bibnamefont
  {Neto}},\ }\href {\doibase 10.1126/science.aac9439} {\bibfield  {journal}
  {\bibinfo  {journal} {Science}\ }\textbf {\bibinfo {volume} {353}},\ \bibinfo
  {pages} {aac9439} (\bibinfo {year} {2016})}\BibitemShut {NoStop}%
\bibitem [{\citenamefont {Cao}\ \emph {et~al.}(2018{\natexlab{a}})\citenamefont
  {Cao}, \citenamefont {Fatemi}, \citenamefont {Fang}, \citenamefont
  {Watanabe}, \citenamefont {Taniguchi}, \citenamefont {Kaxiras},\ and\
  \citenamefont {Jarillo-Herrero}}]{yuancao2018}%
  \BibitemOpen
  \bibfield  {author} {\bibinfo {author} {\bibfnamefont {Y.}~\bibnamefont
  {Cao}}, \bibinfo {author} {\bibfnamefont {V.}~\bibnamefont {Fatemi}},
  \bibinfo {author} {\bibfnamefont {S.}~\bibnamefont {Fang}}, \bibinfo {author}
  {\bibfnamefont {K.}~\bibnamefont {Watanabe}}, \bibinfo {author}
  {\bibfnamefont {T.}~\bibnamefont {Taniguchi}}, \bibinfo {author}
  {\bibfnamefont {E.}~\bibnamefont {Kaxiras}}, \ and\ \bibinfo {author}
  {\bibfnamefont {P.}~\bibnamefont {Jarillo-Herrero}},\ }\href {\doibase
  10.1038/nature26160} {\bibfield  {journal} {\bibinfo  {journal} {Nature}\
  }\textbf {\bibinfo {volume} {556}},\ \bibinfo {pages} {43} (\bibinfo {year}
  {2018}{\natexlab{a}})}\BibitemShut {NoStop}%
\bibitem [{\citenamefont {Yankowitz}\ \emph {et~al.}(2019)\citenamefont
  {Yankowitz}, \citenamefont {Chen}, \citenamefont {Polshyn}, \citenamefont
  {Zhang}, \citenamefont {Watanabe}, \citenamefont {Taniguchi}, \citenamefont
  {Graf}, \citenamefont {Young},\ and\ \citenamefont
  {Dean}}]{Yankowitz_TBG_2019}%
  \BibitemOpen
  \bibfield  {author} {\bibinfo {author} {\bibfnamefont {M.}~\bibnamefont
  {Yankowitz}}, \bibinfo {author} {\bibfnamefont {S.}~\bibnamefont {Chen}},
  \bibinfo {author} {\bibfnamefont {H.}~\bibnamefont {Polshyn}}, \bibinfo
  {author} {\bibfnamefont {Y.}~\bibnamefont {Zhang}}, \bibinfo {author}
  {\bibfnamefont {K.}~\bibnamefont {Watanabe}}, \bibinfo {author}
  {\bibfnamefont {T.}~\bibnamefont {Taniguchi}}, \bibinfo {author}
  {\bibfnamefont {D.}~\bibnamefont {Graf}}, \bibinfo {author} {\bibfnamefont
  {A.}~\bibnamefont {Young}}, \ and\ \bibinfo {author} {\bibfnamefont
  {C.}~\bibnamefont {Dean}},\ }\href {\doibase 10.1126/science.aav1910}
  {\bibfield  {journal} {\bibinfo  {journal} {Science}\ }\textbf {\bibinfo
  {volume} {363}},\ \bibinfo {pages} {eaav1910} (\bibinfo {year}
  {2019})}\BibitemShut {NoStop}%
\bibitem [{\citenamefont {Lu}\ \emph {et~al.}(2019)\citenamefont {Lu},
  \citenamefont {Stepanov}, \citenamefont {Yang}, \citenamefont {Xie},
  \citenamefont {Aamir}, \citenamefont {Das}, \citenamefont {Urgell},
  \citenamefont {Watanabe}, \citenamefont {Taniguchi}, \citenamefont {Zhang},
  \citenamefont {Bachtold}, \citenamefont {MacDonald},\ and\ \citenamefont
  {Efetov}}]{Lu_TBG_2019}%
  \BibitemOpen
  \bibfield  {author} {\bibinfo {author} {\bibfnamefont {X.}~\bibnamefont
  {Lu}}, \bibinfo {author} {\bibfnamefont {P.}~\bibnamefont {Stepanov}},
  \bibinfo {author} {\bibfnamefont {W.}~\bibnamefont {Yang}}, \bibinfo {author}
  {\bibfnamefont {M.}~\bibnamefont {Xie}}, \bibinfo {author} {\bibfnamefont
  {M.~A.}\ \bibnamefont {Aamir}}, \bibinfo {author} {\bibfnamefont
  {I.}~\bibnamefont {Das}}, \bibinfo {author} {\bibfnamefont {C.}~\bibnamefont
  {Urgell}}, \bibinfo {author} {\bibfnamefont {K.}~\bibnamefont {Watanabe}},
  \bibinfo {author} {\bibfnamefont {T.}~\bibnamefont {Taniguchi}}, \bibinfo
  {author} {\bibfnamefont {G.}~\bibnamefont {Zhang}}, \bibinfo {author}
  {\bibfnamefont {A.}~\bibnamefont {Bachtold}}, \bibinfo {author}
  {\bibfnamefont {A.~H.}\ \bibnamefont {MacDonald}}, \ and\ \bibinfo {author}
  {\bibfnamefont {D.~K.}\ \bibnamefont {Efetov}},\ }\href
  {https://EconPapers.repec.org/RePEc:nat:nature:v:574:y:2019:i:7780:d:10.1038_s41586-019-1695-0}
  {\bibfield  {journal} {\bibinfo  {journal} {Nature}\ }\textbf {\bibinfo
  {volume} {574}},\ \bibinfo {pages} {653} (\bibinfo {year}
  {2019})}\BibitemShut {NoStop}%
\bibitem [{\citenamefont {Saito}\ \emph {et~al.}(2020)\citenamefont {Saito},
  \citenamefont {Ge}, \citenamefont {Watanabe}, \citenamefont {Taniguchi},\
  and\ \citenamefont {Young}}]{Saito_TBG_2020}%
  \BibitemOpen
  \bibfield  {author} {\bibinfo {author} {\bibfnamefont {Y.}~\bibnamefont
  {Saito}}, \bibinfo {author} {\bibfnamefont {J.}~\bibnamefont {Ge}}, \bibinfo
  {author} {\bibfnamefont {K.}~\bibnamefont {Watanabe}}, \bibinfo {author}
  {\bibfnamefont {T.}~\bibnamefont {Taniguchi}}, \ and\ \bibinfo {author}
  {\bibfnamefont {A.}~\bibnamefont {Young}},\ }\href {\doibase
  10.1038/s41567-020-0928-3} {\bibfield  {journal} {\bibinfo  {journal} {Nature
  Physics}\ }\textbf {\bibinfo {volume} {16}} (\bibinfo {year} {2020}),\
  10.1038/s41567-020-0928-3}\BibitemShut {NoStop}%
\bibitem [{\citenamefont {Chen}\ \emph
  {et~al.}(2019{\natexlab{a}})\citenamefont {Chen}, \citenamefont {Jiang},
  \citenamefont {Wu}, \citenamefont {Lyu}, \citenamefont {Li}, \citenamefont
  {Chittari}, \citenamefont {Watanabe}, \citenamefont {Taniguchi},
  \citenamefont {Shi}, \citenamefont {Jung}, \citenamefont {Zhang},\ and\
  \citenamefont {Wang}}]{Chen_ABC1_2019}%
  \BibitemOpen
  \bibfield  {author} {\bibinfo {author} {\bibfnamefont {G.}~\bibnamefont
  {Chen}}, \bibinfo {author} {\bibfnamefont {L.}~\bibnamefont {Jiang}},
  \bibinfo {author} {\bibfnamefont {S.}~\bibnamefont {Wu}}, \bibinfo {author}
  {\bibfnamefont {B.}~\bibnamefont {Lyu}}, \bibinfo {author} {\bibfnamefont
  {H.}~\bibnamefont {Li}}, \bibinfo {author} {\bibfnamefont {B.~L.}\
  \bibnamefont {Chittari}}, \bibinfo {author} {\bibfnamefont {K.}~\bibnamefont
  {Watanabe}}, \bibinfo {author} {\bibfnamefont {T.}~\bibnamefont {Taniguchi}},
  \bibinfo {author} {\bibfnamefont {Z.}~\bibnamefont {Shi}}, \bibinfo {author}
  {\bibfnamefont {J.}~\bibnamefont {Jung}}, \bibinfo {author} {\bibfnamefont
  {Y.}~\bibnamefont {Zhang}}, \ and\ \bibinfo {author} {\bibfnamefont
  {F.}~\bibnamefont {Wang}},\ }\href {\doibase 10.1038/s41567-018-0387-2}
  {\bibfield  {journal} {\bibinfo  {journal} {Nature Physics}\ }\textbf
  {\bibinfo {volume} {15}} (\bibinfo {year} {2019}{\natexlab{a}}),\
  10.1038/s41567-018-0387-2}\BibitemShut {NoStop}%
\bibitem [{\citenamefont {Chen}\ \emph
  {et~al.}(2019{\natexlab{b}})\citenamefont {Chen}, \citenamefont {Sharpe},
  \citenamefont {Gallagher}, \citenamefont {Rosen}, \citenamefont {Fox},
  \citenamefont {Jiang}, \citenamefont {Lyu}, \citenamefont {Li}, \citenamefont
  {Watanabe}, \citenamefont {Taniguchi}, \citenamefont {Jung}, \citenamefont
  {Shi}, \citenamefont {Goldhaber-Gordon}, \citenamefont {Zhang},\ and\
  \citenamefont {Wang}}]{Chen_ABC2_2019}%
  \BibitemOpen
  \bibfield  {author} {\bibinfo {author} {\bibfnamefont {G.}~\bibnamefont
  {Chen}}, \bibinfo {author} {\bibfnamefont {A.}~\bibnamefont {Sharpe}},
  \bibinfo {author} {\bibfnamefont {P.}~\bibnamefont {Gallagher}}, \bibinfo
  {author} {\bibfnamefont {I.}~\bibnamefont {Rosen}}, \bibinfo {author}
  {\bibfnamefont {E.}~\bibnamefont {Fox}}, \bibinfo {author} {\bibfnamefont
  {L.}~\bibnamefont {Jiang}}, \bibinfo {author} {\bibfnamefont
  {B.}~\bibnamefont {Lyu}}, \bibinfo {author} {\bibfnamefont {H.}~\bibnamefont
  {Li}}, \bibinfo {author} {\bibfnamefont {K.}~\bibnamefont {Watanabe}},
  \bibinfo {author} {\bibfnamefont {T.}~\bibnamefont {Taniguchi}}, \bibinfo
  {author} {\bibfnamefont {J.}~\bibnamefont {Jung}}, \bibinfo {author}
  {\bibfnamefont {Z.}~\bibnamefont {Shi}}, \bibinfo {author} {\bibfnamefont
  {D.}~\bibnamefont {Goldhaber-Gordon}}, \bibinfo {author} {\bibfnamefont
  {Y.}~\bibnamefont {Zhang}}, \ and\ \bibinfo {author} {\bibfnamefont
  {F.}~\bibnamefont {Wang}},\ }\href {\doibase 10.1038/s41586-019-1393-y}
  {\bibfield  {journal} {\bibinfo  {journal} {Nature}\ }\textbf {\bibinfo
  {volume} {572}},\ \bibinfo {pages} {1} (\bibinfo {year}
  {2019}{\natexlab{b}})}\BibitemShut {NoStop}%
\bibitem [{\citenamefont {Park}\ \emph {et~al.}(2021)\citenamefont {Park},
  \citenamefont {Cao}, \citenamefont {Watanabe}, \citenamefont {Taniguchi},\
  and\ \citenamefont {Jarillo-Herrero}}]{Cao_TTG_2021}%
  \BibitemOpen
  \bibfield  {author} {\bibinfo {author} {\bibfnamefont {J.}~\bibnamefont
  {Park}}, \bibinfo {author} {\bibfnamefont {Y.}~\bibnamefont {Cao}}, \bibinfo
  {author} {\bibfnamefont {K.}~\bibnamefont {Watanabe}}, \bibinfo {author}
  {\bibfnamefont {T.}~\bibnamefont {Taniguchi}}, \ and\ \bibinfo {author}
  {\bibfnamefont {P.}~\bibnamefont {Jarillo-Herrero}},\ }\href {\doibase
  10.1038/s41586-021-03192-0} {\bibfield  {journal} {\bibinfo  {journal}
  {Nature}\ }\textbf {\bibinfo {volume} {590}},\ \bibinfo {pages} {1} (\bibinfo
  {year} {2021})}\BibitemShut {NoStop}%
\bibitem [{\citenamefont {Zhang}\ \emph {et~al.}(2021)\citenamefont {Zhang},
  \citenamefont {Tsai}, \citenamefont {Zhu}, \citenamefont {Ren}, \citenamefont
  {Luo}, \citenamefont {Carr}, \citenamefont {Luskin}, \citenamefont
  {Kaxiras},\ and\ \citenamefont {Wang}}]{Tsai_TTG_2021}%
  \BibitemOpen
  \bibfield  {author} {\bibinfo {author} {\bibfnamefont {X.}~\bibnamefont
  {Zhang}}, \bibinfo {author} {\bibfnamefont {K.-T.}\ \bibnamefont {Tsai}},
  \bibinfo {author} {\bibfnamefont {Z.}~\bibnamefont {Zhu}}, \bibinfo {author}
  {\bibfnamefont {W.}~\bibnamefont {Ren}}, \bibinfo {author} {\bibfnamefont
  {Y.}~\bibnamefont {Luo}}, \bibinfo {author} {\bibfnamefont {S.}~\bibnamefont
  {Carr}}, \bibinfo {author} {\bibfnamefont {M.}~\bibnamefont {Luskin}},
  \bibinfo {author} {\bibfnamefont {E.}~\bibnamefont {Kaxiras}}, \ and\
  \bibinfo {author} {\bibfnamefont {K.}~\bibnamefont {Wang}},\ }\href {\doibase
  10.1103/PhysRevLett.127.166802} {\bibfield  {journal} {\bibinfo  {journal}
  {Phys. Rev. Lett.}\ }\textbf {\bibinfo {volume} {127}},\ \bibinfo {pages}
  {166802} (\bibinfo {year} {2021})}\BibitemShut {NoStop}%
\bibitem [{\citenamefont {Park}\ \emph
  {et~al.}(2022{\natexlab{a}})\citenamefont {Park}, \citenamefont {Cao},
  \citenamefont {Xia}, \citenamefont {Sun}, \citenamefont {Watanabe},
  \citenamefont {Taniguchi},\ and\ \citenamefont
  {Jarillo-Herrero}}]{Cao_mutiLG_2022}%
  \BibitemOpen
  \bibfield  {author} {\bibinfo {author} {\bibfnamefont {J.}~\bibnamefont
  {Park}}, \bibinfo {author} {\bibfnamefont {Y.}~\bibnamefont {Cao}}, \bibinfo
  {author} {\bibfnamefont {L.-Q.}\ \bibnamefont {Xia}}, \bibinfo {author}
  {\bibfnamefont {S.}~\bibnamefont {Sun}}, \bibinfo {author} {\bibfnamefont
  {K.}~\bibnamefont {Watanabe}}, \bibinfo {author} {\bibfnamefont
  {T.}~\bibnamefont {Taniguchi}}, \ and\ \bibinfo {author} {\bibfnamefont
  {P.}~\bibnamefont {Jarillo-Herrero}},\ }\href {\doibase
  10.1038/s41563-022-01287-1} {\bibfield  {journal} {\bibinfo  {journal}
  {Nature Materials}\ } (\bibinfo {year} {2022}{\natexlab{a}}),\
  10.1038/s41563-022-01287-1}\BibitemShut {NoStop}%
\bibitem [{\citenamefont {Po}\ \emph {et~al.}(2018)\citenamefont {Po},
  \citenamefont {Zou}, \citenamefont {Vishwanath},\ and\ \citenamefont
  {Senthil}}]{PhysRevX.8.031089}%
  \BibitemOpen
  \bibfield  {author} {\bibinfo {author} {\bibfnamefont {H.~C.}\ \bibnamefont
  {Po}}, \bibinfo {author} {\bibfnamefont {L.}~\bibnamefont {Zou}}, \bibinfo
  {author} {\bibfnamefont {A.}~\bibnamefont {Vishwanath}}, \ and\ \bibinfo
  {author} {\bibfnamefont {T.}~\bibnamefont {Senthil}},\ }\href {\doibase
  10.1103/PhysRevX.8.031089} {\bibfield  {journal} {\bibinfo  {journal} {Phys.
  Rev. X}\ }\textbf {\bibinfo {volume} {8}},\ \bibinfo {pages} {031089}
  (\bibinfo {year} {2018})}\BibitemShut {NoStop}%
\bibitem [{\citenamefont {Isobe}\ \emph {et~al.}(2018)\citenamefont {Isobe},
  \citenamefont {Yuan},\ and\ \citenamefont {Fu}}]{PhysRevX.8.041041}%
  \BibitemOpen
  \bibfield  {author} {\bibinfo {author} {\bibfnamefont {H.}~\bibnamefont
  {Isobe}}, \bibinfo {author} {\bibfnamefont {N.~F.~Q.}\ \bibnamefont {Yuan}},
  \ and\ \bibinfo {author} {\bibfnamefont {L.}~\bibnamefont {Fu}},\ }\href
  {\doibase 10.1103/PhysRevX.8.041041} {\bibfield  {journal} {\bibinfo
  {journal} {Phys. Rev. X}\ }\textbf {\bibinfo {volume} {8}},\ \bibinfo {pages}
  {041041} (\bibinfo {year} {2018})}\BibitemShut {NoStop}%
\bibitem [{\citenamefont {Wu}\ \emph {et~al.}(2018)\citenamefont {Wu},
  \citenamefont {MacDonald},\ and\ \citenamefont
  {Martin}}]{PhysRevLett.121.257001}%
  \BibitemOpen
  \bibfield  {author} {\bibinfo {author} {\bibfnamefont {F.}~\bibnamefont
  {Wu}}, \bibinfo {author} {\bibfnamefont {A.~H.}\ \bibnamefont {MacDonald}}, \
  and\ \bibinfo {author} {\bibfnamefont {I.}~\bibnamefont {Martin}},\ }\href
  {\doibase 10.1103/PhysRevLett.121.257001} {\bibfield  {journal} {\bibinfo
  {journal} {Phys. Rev. Lett.}\ }\textbf {\bibinfo {volume} {121}},\ \bibinfo
  {pages} {257001} (\bibinfo {year} {2018})}\BibitemShut {NoStop}%
\bibitem [{\citenamefont {Xu}\ and\ \citenamefont
  {Balents}(2018)}]{PhysRevLett.121.087001}%
  \BibitemOpen
  \bibfield  {author} {\bibinfo {author} {\bibfnamefont {C.}~\bibnamefont
  {Xu}}\ and\ \bibinfo {author} {\bibfnamefont {L.}~\bibnamefont {Balents}},\
  }\href {\doibase 10.1103/PhysRevLett.121.087001} {\bibfield  {journal}
  {\bibinfo  {journal} {Phys. Rev. Lett.}\ }\textbf {\bibinfo {volume} {121}},\
  \bibinfo {pages} {087001} (\bibinfo {year} {2018})}\BibitemShut {NoStop}%
\bibitem [{\citenamefont {Xie}\ \emph {et~al.}(2020)\citenamefont {Xie},
  \citenamefont {Song}, \citenamefont {Lian},\ and\ \citenamefont
  {Bernevig}}]{PhysRevLett.124.167002}%
  \BibitemOpen
  \bibfield  {author} {\bibinfo {author} {\bibfnamefont {F.}~\bibnamefont
  {Xie}}, \bibinfo {author} {\bibfnamefont {Z.}~\bibnamefont {Song}}, \bibinfo
  {author} {\bibfnamefont {B.}~\bibnamefont {Lian}}, \ and\ \bibinfo {author}
  {\bibfnamefont {B.~A.}\ \bibnamefont {Bernevig}},\ }\href {\doibase
  10.1103/PhysRevLett.124.167002} {\bibfield  {journal} {\bibinfo  {journal}
  {Phys. Rev. Lett.}\ }\textbf {\bibinfo {volume} {124}},\ \bibinfo {pages}
  {167002} (\bibinfo {year} {2020})}\BibitemShut {NoStop}%
\bibitem [{\citenamefont {Song}\ and\ \citenamefont
  {Bernevig}(2022)}]{PhysRevLett.129.047601}%
  \BibitemOpen
  \bibfield  {author} {\bibinfo {author} {\bibfnamefont {Z.-D.}\ \bibnamefont
  {Song}}\ and\ \bibinfo {author} {\bibfnamefont {B.~A.}\ \bibnamefont
  {Bernevig}},\ }\href {\doibase 10.1103/PhysRevLett.129.047601} {\bibfield
  {journal} {\bibinfo  {journal} {Phys. Rev. Lett.}\ }\textbf {\bibinfo
  {volume} {129}},\ \bibinfo {pages} {047601} (\bibinfo {year}
  {2022})}\BibitemShut {NoStop}%
\bibitem [{\citenamefont {Liu}\ \emph {et~al.}(2018)\citenamefont {Liu},
  \citenamefont {Zhang}, \citenamefont {Chen},\ and\ \citenamefont
  {Yang}}]{PhysRevLett.121.217001}%
  \BibitemOpen
  \bibfield  {author} {\bibinfo {author} {\bibfnamefont {C.-C.}\ \bibnamefont
  {Liu}}, \bibinfo {author} {\bibfnamefont {L.-D.}\ \bibnamefont {Zhang}},
  \bibinfo {author} {\bibfnamefont {W.-Q.}\ \bibnamefont {Chen}}, \ and\
  \bibinfo {author} {\bibfnamefont {F.}~\bibnamefont {Yang}},\ }\href {\doibase
  10.1103/PhysRevLett.121.217001} {\bibfield  {journal} {\bibinfo  {journal}
  {Phys. Rev. Lett.}\ }\textbf {\bibinfo {volume} {121}},\ \bibinfo {pages}
  {217001} (\bibinfo {year} {2018})}\BibitemShut {NoStop}%
\bibitem [{\citenamefont {Roy}\ and\ \citenamefont {Juri\ifmmode \check{c}\else
  \v{c}\fi{}i\ifmmode~\acute{c}\else \'{c}\fi{}}(2019)}]{PhysRevB.99.121407}%
  \BibitemOpen
  \bibfield  {author} {\bibinfo {author} {\bibfnamefont {B.}~\bibnamefont
  {Roy}}\ and\ \bibinfo {author} {\bibfnamefont {V.}~\bibnamefont {Juri\ifmmode
  \check{c}\else \v{c}\fi{}i\ifmmode~\acute{c}\else \'{c}\fi{}}},\ }\href
  {\doibase 10.1103/PhysRevB.99.121407} {\bibfield  {journal} {\bibinfo
  {journal} {Phys. Rev. B}\ }\textbf {\bibinfo {volume} {99}},\ \bibinfo
  {pages} {121407} (\bibinfo {year} {2019})}\BibitemShut {NoStop}%
\bibitem [{\citenamefont {Cao}\ \emph {et~al.}(2018{\natexlab{b}})\citenamefont
  {Cao}, \citenamefont {Fatemi}, \citenamefont {Demir}, \citenamefont {Fang},
  \citenamefont {Tomarken}, \citenamefont {Luo}, \citenamefont
  {Sanchez-Yamagishi}, \citenamefont {Watanabe}, \citenamefont {Taniguchi},
  \citenamefont {Kaxiras}, \citenamefont {Ashoori},\ and\ \citenamefont
  {Jarillo-Herrero}}]{cao2018correlated}%
  \BibitemOpen
  \bibfield  {author} {\bibinfo {author} {\bibfnamefont {Y.}~\bibnamefont
  {Cao}}, \bibinfo {author} {\bibfnamefont {V.}~\bibnamefont {Fatemi}},
  \bibinfo {author} {\bibfnamefont {A.}~\bibnamefont {Demir}}, \bibinfo
  {author} {\bibfnamefont {S.}~\bibnamefont {Fang}}, \bibinfo {author}
  {\bibfnamefont {S.~L.}\ \bibnamefont {Tomarken}}, \bibinfo {author}
  {\bibfnamefont {J.~Y.}\ \bibnamefont {Luo}}, \bibinfo {author} {\bibfnamefont
  {J.~D.}\ \bibnamefont {Sanchez-Yamagishi}}, \bibinfo {author} {\bibfnamefont
  {K.}~\bibnamefont {Watanabe}}, \bibinfo {author} {\bibfnamefont
  {T.}~\bibnamefont {Taniguchi}}, \bibinfo {author} {\bibfnamefont
  {E.}~\bibnamefont {Kaxiras}}, \bibinfo {author} {\bibfnamefont {R.~C.}\
  \bibnamefont {Ashoori}}, \ and\ \bibinfo {author} {\bibfnamefont
  {P.}~\bibnamefont {Jarillo-Herrero}},\ }\href {\doibase 10.1038/nature26154}
  {\bibfield  {journal} {\bibinfo  {journal} {Nature}\ }\textbf {\bibinfo
  {volume} {556}},\ \bibinfo {pages} {80} (\bibinfo {year}
  {2018}{\natexlab{b}})}\BibitemShut {NoStop}%
\bibitem [{\citenamefont {Cao}\ \emph {et~al.}(2020)\citenamefont {Cao},
  \citenamefont {Chowdhury}, \citenamefont {Rodan-Legrain}, \citenamefont
  {Rubies-Bigorda}, \citenamefont {Watanabe}, \citenamefont {Taniguchi},
  \citenamefont {Senthil},\ and\ \citenamefont
  {Jarillo-Herrero}}]{Cao_TBG_strange_2020}%
  \BibitemOpen
  \bibfield  {author} {\bibinfo {author} {\bibfnamefont {Y.}~\bibnamefont
  {Cao}}, \bibinfo {author} {\bibfnamefont {D.}~\bibnamefont {Chowdhury}},
  \bibinfo {author} {\bibfnamefont {D.}~\bibnamefont {Rodan-Legrain}}, \bibinfo
  {author} {\bibfnamefont {O.}~\bibnamefont {Rubies-Bigorda}}, \bibinfo
  {author} {\bibfnamefont {K.}~\bibnamefont {Watanabe}}, \bibinfo {author}
  {\bibfnamefont {T.}~\bibnamefont {Taniguchi}}, \bibinfo {author}
  {\bibfnamefont {T.}~\bibnamefont {Senthil}}, \ and\ \bibinfo {author}
  {\bibfnamefont {P.}~\bibnamefont {Jarillo-Herrero}},\ }\href {\doibase
  10.1103/PhysRevLett.124.076801} {\bibfield  {journal} {\bibinfo  {journal}
  {Phys. Rev. Lett.}\ }\textbf {\bibinfo {volume} {124}},\ \bibinfo {pages}
  {076801} (\bibinfo {year} {2020})}\BibitemShut {NoStop}%
\bibitem [{\citenamefont {Liu}\ \emph {et~al.}(2021)\citenamefont {Liu},
  \citenamefont {Wang}, \citenamefont {Watanabe}, \citenamefont {Taniguchi},
  \citenamefont {Vafek},\ and\ \citenamefont {Li}}]{liu2021tuning}%
  \BibitemOpen
  \bibfield  {author} {\bibinfo {author} {\bibfnamefont {X.}~\bibnamefont
  {Liu}}, \bibinfo {author} {\bibfnamefont {Z.}~\bibnamefont {Wang}}, \bibinfo
  {author} {\bibfnamefont {K.}~\bibnamefont {Watanabe}}, \bibinfo {author}
  {\bibfnamefont {T.}~\bibnamefont {Taniguchi}}, \bibinfo {author}
  {\bibfnamefont {O.}~\bibnamefont {Vafek}}, \ and\ \bibinfo {author}
  {\bibfnamefont {J.}~\bibnamefont {Li}},\ }\href@noop {} {\bibfield  {journal}
  {\bibinfo  {journal} {Science}\ }\textbf {\bibinfo {volume} {371}},\ \bibinfo
  {pages} {1261} (\bibinfo {year} {2021})}\BibitemShut {NoStop}%
\bibitem [{\citenamefont {Xie}\ and\ \citenamefont
  {MacDonald}(2020)}]{PhysRevLett.124.097601}%
  \BibitemOpen
  \bibfield  {author} {\bibinfo {author} {\bibfnamefont {M.}~\bibnamefont
  {Xie}}\ and\ \bibinfo {author} {\bibfnamefont {A.~H.}\ \bibnamefont
  {MacDonald}},\ }\href {\doibase 10.1103/PhysRevLett.124.097601} {\bibfield
  {journal} {\bibinfo  {journal} {Phys. Rev. Lett.}\ }\textbf {\bibinfo
  {volume} {124}},\ \bibinfo {pages} {097601} (\bibinfo {year}
  {2020})}\BibitemShut {NoStop}%
\bibitem [{\citenamefont {Koshino}\ \emph {et~al.}(2018)\citenamefont
  {Koshino}, \citenamefont {Yuan}, \citenamefont {Koretsune}, \citenamefont
  {Ochi}, \citenamefont {Kuroki},\ and\ \citenamefont
  {Fu}}]{PhysRevX.8.031087}%
  \BibitemOpen
  \bibfield  {author} {\bibinfo {author} {\bibfnamefont {M.}~\bibnamefont
  {Koshino}}, \bibinfo {author} {\bibfnamefont {N.~F.~Q.}\ \bibnamefont
  {Yuan}}, \bibinfo {author} {\bibfnamefont {T.}~\bibnamefont {Koretsune}},
  \bibinfo {author} {\bibfnamefont {M.}~\bibnamefont {Ochi}}, \bibinfo {author}
  {\bibfnamefont {K.}~\bibnamefont {Kuroki}}, \ and\ \bibinfo {author}
  {\bibfnamefont {L.}~\bibnamefont {Fu}},\ }\href {\doibase
  10.1103/PhysRevX.8.031087} {\bibfield  {journal} {\bibinfo  {journal} {Phys.
  Rev. X}\ }\textbf {\bibinfo {volume} {8}},\ \bibinfo {pages} {031087}
  (\bibinfo {year} {2018})}\BibitemShut {NoStop}%
\bibitem [{\citenamefont {Dodaro}\ \emph {et~al.}(2018)\citenamefont {Dodaro},
  \citenamefont {Kivelson}, \citenamefont {Schattner}, \citenamefont {Sun},\
  and\ \citenamefont {Wang}}]{PhysRevB.98.075154}%
  \BibitemOpen
  \bibfield  {author} {\bibinfo {author} {\bibfnamefont {J.~F.}\ \bibnamefont
  {Dodaro}}, \bibinfo {author} {\bibfnamefont {S.~A.}\ \bibnamefont
  {Kivelson}}, \bibinfo {author} {\bibfnamefont {Y.}~\bibnamefont {Schattner}},
  \bibinfo {author} {\bibfnamefont {X.~Q.}\ \bibnamefont {Sun}}, \ and\
  \bibinfo {author} {\bibfnamefont {C.}~\bibnamefont {Wang}},\ }\href {\doibase
  10.1103/PhysRevB.98.075154} {\bibfield  {journal} {\bibinfo  {journal} {Phys.
  Rev. B}\ }\textbf {\bibinfo {volume} {98}},\ \bibinfo {pages} {075154}
  (\bibinfo {year} {2018})}\BibitemShut {NoStop}%
\bibitem [{\citenamefont {Yuan}\ and\ \citenamefont
  {Fu}(2018)}]{PhysRevB.98.045103}%
  \BibitemOpen
  \bibfield  {author} {\bibinfo {author} {\bibfnamefont {N.~F.~Q.}\
  \bibnamefont {Yuan}}\ and\ \bibinfo {author} {\bibfnamefont {L.}~\bibnamefont
  {Fu}},\ }\href {\doibase 10.1103/PhysRevB.98.045103} {\bibfield  {journal}
  {\bibinfo  {journal} {Phys. Rev. B}\ }\textbf {\bibinfo {volume} {98}},\
  \bibinfo {pages} {045103} (\bibinfo {year} {2018})}\BibitemShut {NoStop}%
\bibitem [{\citenamefont {Thomson}\ \emph {et~al.}(2018)\citenamefont
  {Thomson}, \citenamefont {Chatterjee}, \citenamefont {Sachdev},\ and\
  \citenamefont {Scheurer}}]{PhysRevB.98.075109}%
  \BibitemOpen
  \bibfield  {author} {\bibinfo {author} {\bibfnamefont {A.}~\bibnamefont
  {Thomson}}, \bibinfo {author} {\bibfnamefont {S.}~\bibnamefont {Chatterjee}},
  \bibinfo {author} {\bibfnamefont {S.}~\bibnamefont {Sachdev}}, \ and\
  \bibinfo {author} {\bibfnamefont {M.~S.}\ \bibnamefont {Scheurer}},\ }\href
  {\doibase 10.1103/PhysRevB.98.075109} {\bibfield  {journal} {\bibinfo
  {journal} {Phys. Rev. B}\ }\textbf {\bibinfo {volume} {98}},\ \bibinfo
  {pages} {075109} (\bibinfo {year} {2018})}\BibitemShut {NoStop}%
\bibitem [{\citenamefont {Pizarro}\ \emph {et~al.}(2019)\citenamefont
  {Pizarro}, \citenamefont {Calder{\'o}n},\ and\ \citenamefont
  {Bascones}}]{pizarro2019nature}%
  \BibitemOpen
  \bibfield  {author} {\bibinfo {author} {\bibfnamefont {J.~M.}\ \bibnamefont
  {Pizarro}}, \bibinfo {author} {\bibfnamefont {M.}~\bibnamefont
  {Calder{\'o}n}}, \ and\ \bibinfo {author} {\bibfnamefont {E.}~\bibnamefont
  {Bascones}},\ }\href@noop {} {\bibfield  {journal} {\bibinfo  {journal}
  {Journal of Physics Communications}\ }\textbf {\bibinfo {volume} {3}},\
  \bibinfo {pages} {035024} (\bibinfo {year} {2019})}\BibitemShut {NoStop}%
\bibitem [{\citenamefont {Park}\ \emph
  {et~al.}(2022{\natexlab{b}})\citenamefont {Park}, \citenamefont {Cao},
  \citenamefont {Xia}, \citenamefont {Sun}, \citenamefont {Watanabe},
  \citenamefont {Taniguchi},\ and\ \citenamefont
  {Jarillo-Herrero}}]{parkcao2022}%
  \BibitemOpen
  \bibfield  {author} {\bibinfo {author} {\bibfnamefont {J.~M.}\ \bibnamefont
  {Park}}, \bibinfo {author} {\bibfnamefont {Y.}~\bibnamefont {Cao}}, \bibinfo
  {author} {\bibfnamefont {L.-Q.}\ \bibnamefont {Xia}}, \bibinfo {author}
  {\bibfnamefont {S.}~\bibnamefont {Sun}}, \bibinfo {author} {\bibfnamefont
  {K.}~\bibnamefont {Watanabe}}, \bibinfo {author} {\bibfnamefont
  {T.}~\bibnamefont {Taniguchi}}, \ and\ \bibinfo {author} {\bibfnamefont
  {P.}~\bibnamefont {Jarillo-Herrero}},\ }\href {\doibase
  10.1038/s41563-022-01287-1} {\bibfield  {journal} {\bibinfo  {journal}
  {Nature Materials}\ }\textbf {\bibinfo {volume} {21}},\ \bibinfo {pages}
  {877} (\bibinfo {year} {2022}{\natexlab{b}})}\BibitemShut {NoStop}%
\bibitem [{\citenamefont {Wilson}\ \emph {et~al.}(2020)\citenamefont {Wilson},
  \citenamefont {Fu}, \citenamefont {Das~Sarma},\ and\ \citenamefont
  {Pixley}}]{PhysRevResearch.2.023325}%
  \BibitemOpen
  \bibfield  {author} {\bibinfo {author} {\bibfnamefont {J.~H.}\ \bibnamefont
  {Wilson}}, \bibinfo {author} {\bibfnamefont {Y.}~\bibnamefont {Fu}}, \bibinfo
  {author} {\bibfnamefont {S.}~\bibnamefont {Das~Sarma}}, \ and\ \bibinfo
  {author} {\bibfnamefont {J.~H.}\ \bibnamefont {Pixley}},\ }\href {\doibase
  10.1103/PhysRevResearch.2.023325} {\bibfield  {journal} {\bibinfo  {journal}
  {Phys. Rev. Res.}\ }\textbf {\bibinfo {volume} {2}},\ \bibinfo {pages}
  {023325} (\bibinfo {year} {2020})}\BibitemShut {NoStop}%
\bibitem [{\citenamefont {Bi}\ \emph {et~al.}(2019)\citenamefont {Bi},
  \citenamefont {Yuan},\ and\ \citenamefont {Fu}}]{PhysRevB.100.035448}%
  \BibitemOpen
  \bibfield  {author} {\bibinfo {author} {\bibfnamefont {Z.}~\bibnamefont
  {Bi}}, \bibinfo {author} {\bibfnamefont {N.~F.~Q.}\ \bibnamefont {Yuan}}, \
  and\ \bibinfo {author} {\bibfnamefont {L.}~\bibnamefont {Fu}},\ }\href
  {\doibase 10.1103/PhysRevB.100.035448} {\bibfield  {journal} {\bibinfo
  {journal} {Phys. Rev. B}\ }\textbf {\bibinfo {volume} {100}},\ \bibinfo
  {pages} {035448} (\bibinfo {year} {2019})}\BibitemShut {NoStop}%
\bibitem [{\citenamefont {Choi}\ and\ \citenamefont
  {Choi}(2019)}]{choi2019intrinsic}%
  \BibitemOpen
  \bibfield  {author} {\bibinfo {author} {\bibfnamefont {Y.~W.}\ \bibnamefont
  {Choi}}\ and\ \bibinfo {author} {\bibfnamefont {H.~J.}\ \bibnamefont
  {Choi}},\ }\href@noop {} {\bibfield  {journal} {\bibinfo  {journal} {Physical
  Review B}\ }\textbf {\bibinfo {volume} {100}},\ \bibinfo {pages} {201402}
  (\bibinfo {year} {2019})}\BibitemShut {NoStop}%
\bibitem [{\citenamefont {Liu}\ \emph {et~al.}(2020)\citenamefont {Liu},
  \citenamefont {Su}, \citenamefont {Zhou}, \citenamefont {Yin}, \citenamefont
  {Yan}, \citenamefont {Li}, \citenamefont {Yan}, \citenamefont {Han},
  \citenamefont {Fu}, \citenamefont {Zhang} \emph {et~al.}}]{liu2020tunable}%
  \BibitemOpen
  \bibfield  {author} {\bibinfo {author} {\bibfnamefont {Y.-W.}\ \bibnamefont
  {Liu}}, \bibinfo {author} {\bibfnamefont {Y.}~\bibnamefont {Su}}, \bibinfo
  {author} {\bibfnamefont {X.-F.}\ \bibnamefont {Zhou}}, \bibinfo {author}
  {\bibfnamefont {L.-J.}\ \bibnamefont {Yin}}, \bibinfo {author} {\bibfnamefont
  {C.}~\bibnamefont {Yan}}, \bibinfo {author} {\bibfnamefont {S.-Y.}\
  \bibnamefont {Li}}, \bibinfo {author} {\bibfnamefont {W.}~\bibnamefont
  {Yan}}, \bibinfo {author} {\bibfnamefont {S.}~\bibnamefont {Han}}, \bibinfo
  {author} {\bibfnamefont {Z.-Q.}\ \bibnamefont {Fu}}, \bibinfo {author}
  {\bibfnamefont {Y.}~\bibnamefont {Zhang}},  \emph {et~al.},\ }\href@noop {}
  {\bibfield  {journal} {\bibinfo  {journal} {Physical Review Letters}\
  }\textbf {\bibinfo {volume} {125}},\ \bibinfo {pages} {236102} (\bibinfo
  {year} {2020})}\BibitemShut {NoStop}%
\bibitem [{\citenamefont {Sboychakov}\ \emph {et~al.}(2018)\citenamefont
  {Sboychakov}, \citenamefont {Rozhkov}, \citenamefont {Rakhmanov},\ and\
  \citenamefont {Nori}}]{PhysRevLett.120.266402}%
  \BibitemOpen
  \bibfield  {author} {\bibinfo {author} {\bibfnamefont {A.~O.}\ \bibnamefont
  {Sboychakov}}, \bibinfo {author} {\bibfnamefont {A.~V.}\ \bibnamefont
  {Rozhkov}}, \bibinfo {author} {\bibfnamefont {A.~L.}\ \bibnamefont
  {Rakhmanov}}, \ and\ \bibinfo {author} {\bibfnamefont {F.}~\bibnamefont
  {Nori}},\ }\href {\doibase 10.1103/PhysRevLett.120.266402} {\bibfield
  {journal} {\bibinfo  {journal} {Phys. Rev. Lett.}\ }\textbf {\bibinfo
  {volume} {120}},\ \bibinfo {pages} {266402} (\bibinfo {year}
  {2018})}\BibitemShut {NoStop}%
\bibitem [{\citenamefont {Wolf}\ \emph {et~al.}(2019)\citenamefont {Wolf},
  \citenamefont {Lado}, \citenamefont {Blatter},\ and\ \citenamefont
  {Zilberberg}}]{PhysRevLett.123.096802}%
  \BibitemOpen
  \bibfield  {author} {\bibinfo {author} {\bibfnamefont {T.~M.~R.}\
  \bibnamefont {Wolf}}, \bibinfo {author} {\bibfnamefont {J.~L.}\ \bibnamefont
  {Lado}}, \bibinfo {author} {\bibfnamefont {G.}~\bibnamefont {Blatter}}, \
  and\ \bibinfo {author} {\bibfnamefont {O.}~\bibnamefont {Zilberberg}},\
  }\href {\doibase 10.1103/PhysRevLett.123.096802} {\bibfield  {journal}
  {\bibinfo  {journal} {Phys. Rev. Lett.}\ }\textbf {\bibinfo {volume} {123}},\
  \bibinfo {pages} {096802} (\bibinfo {year} {2019})}\BibitemShut {NoStop}%
\bibitem [{\citenamefont {H{\"u}bener}\ \emph {et~al.}(2021)\citenamefont
  {H{\"u}bener}, \citenamefont {De~Giovannini}, \citenamefont {Sch{\"a}fer},
  \citenamefont {Andberger}, \citenamefont {Ruggenthaler}, \citenamefont
  {Faist},\ and\ \citenamefont {Rubio}}]{hubener2021engineering}%
  \BibitemOpen
  \bibfield  {author} {\bibinfo {author} {\bibfnamefont {H.}~\bibnamefont
  {H{\"u}bener}}, \bibinfo {author} {\bibfnamefont {U.}~\bibnamefont
  {De~Giovannini}}, \bibinfo {author} {\bibfnamefont {C.}~\bibnamefont
  {Sch{\"a}fer}}, \bibinfo {author} {\bibfnamefont {J.}~\bibnamefont
  {Andberger}}, \bibinfo {author} {\bibfnamefont {M.}~\bibnamefont
  {Ruggenthaler}}, \bibinfo {author} {\bibfnamefont {J.}~\bibnamefont {Faist}},
  \ and\ \bibinfo {author} {\bibfnamefont {A.}~\bibnamefont {Rubio}},\
  }\href@noop {} {\bibfield  {journal} {\bibinfo  {journal} {Nature materials}\
  }\textbf {\bibinfo {volume} {20}},\ \bibinfo {pages} {438} (\bibinfo {year}
  {2021})}\BibitemShut {NoStop}%
\bibitem [{\citenamefont {Schlawin}\ \emph {et~al.}(2022)\citenamefont
  {Schlawin}, \citenamefont {Kennes},\ and\ \citenamefont
  {Sentef}}]{schlawin2022cavity}%
  \BibitemOpen
  \bibfield  {author} {\bibinfo {author} {\bibfnamefont {F.}~\bibnamefont
  {Schlawin}}, \bibinfo {author} {\bibfnamefont {D.~M.}\ \bibnamefont
  {Kennes}}, \ and\ \bibinfo {author} {\bibfnamefont {M.~A.}\ \bibnamefont
  {Sentef}},\ }\href@noop {} {\bibfield  {journal} {\bibinfo  {journal}
  {Applied Physics Reviews}\ }\textbf {\bibinfo {volume} {9}},\ \bibinfo
  {pages} {011312} (\bibinfo {year} {2022})}\BibitemShut {NoStop}%
\bibitem [{\citenamefont {Bloch}\ \emph {et~al.}(2022)\citenamefont {Bloch},
  \citenamefont {Cavalleri}, \citenamefont {Galitski}, \citenamefont {Hafezi},\
  and\ \citenamefont {Rubio}}]{bloch2022strongly}%
  \BibitemOpen
  \bibfield  {author} {\bibinfo {author} {\bibfnamefont {J.}~\bibnamefont
  {Bloch}}, \bibinfo {author} {\bibfnamefont {A.}~\bibnamefont {Cavalleri}},
  \bibinfo {author} {\bibfnamefont {V.}~\bibnamefont {Galitski}}, \bibinfo
  {author} {\bibfnamefont {M.}~\bibnamefont {Hafezi}}, \ and\ \bibinfo {author}
  {\bibfnamefont {A.}~\bibnamefont {Rubio}},\ }\href@noop {} {\bibfield
  {journal} {\bibinfo  {journal} {Nature}\ }\textbf {\bibinfo {volume} {606}},\
  \bibinfo {pages} {41} (\bibinfo {year} {2022})}\BibitemShut {NoStop}%
\bibitem [{\citenamefont {Valagiannopoulos}(2022)}]{PhysRevApplied.18.044011}%
  \BibitemOpen
  \bibfield  {author} {\bibinfo {author} {\bibfnamefont {C.}~\bibnamefont
  {Valagiannopoulos}},\ }\href {\doibase 10.1103/PhysRevApplied.18.044011}
  {\bibfield  {journal} {\bibinfo  {journal} {Phys. Rev. Appl.}\ }\textbf
  {\bibinfo {volume} {18}},\ \bibinfo {pages} {044011} (\bibinfo {year}
  {2022})}\BibitemShut {NoStop}%
\bibitem [{\citenamefont {Sch{\"a}fer}\ \emph {et~al.}(2018)\citenamefont
  {Sch{\"a}fer}, \citenamefont {Ruggenthaler},\ and\ \citenamefont
  {Rubio}}]{schafer2018ab}%
  \BibitemOpen
  \bibfield  {author} {\bibinfo {author} {\bibfnamefont {C.}~\bibnamefont
  {Sch{\"a}fer}}, \bibinfo {author} {\bibfnamefont {M.}~\bibnamefont
  {Ruggenthaler}}, \ and\ \bibinfo {author} {\bibfnamefont {A.}~\bibnamefont
  {Rubio}},\ }\href@noop {} {\bibfield  {journal} {\bibinfo  {journal}
  {Physical Review A}\ }\textbf {\bibinfo {volume} {98}},\ \bibinfo {pages}
  {043801} (\bibinfo {year} {2018})}\BibitemShut {NoStop}%
\bibitem [{\citenamefont {Bacciconi}\ \emph {et~al.}(2023)\citenamefont
  {Bacciconi}, \citenamefont {Andolina}, \citenamefont {Chanda}, \citenamefont
  {Chiriac{\`o}}, \citenamefont {Schir{\'o}},\ and\ \citenamefont
  {Dalmonte}}]{bacciconi2023first}%
  \BibitemOpen
  \bibfield  {author} {\bibinfo {author} {\bibfnamefont {Z.}~\bibnamefont
  {Bacciconi}}, \bibinfo {author} {\bibfnamefont {G.~M.}\ \bibnamefont
  {Andolina}}, \bibinfo {author} {\bibfnamefont {T.}~\bibnamefont {Chanda}},
  \bibinfo {author} {\bibfnamefont {G.}~\bibnamefont {Chiriac{\`o}}}, \bibinfo
  {author} {\bibfnamefont {M.}~\bibnamefont {Schir{\'o}}}, \ and\ \bibinfo
  {author} {\bibfnamefont {M.}~\bibnamefont {Dalmonte}},\ }\href@noop {}
  {\bibfield  {journal} {\bibinfo  {journal} {arXiv preprint arXiv:2302.09901}\
  } (\bibinfo {year} {2023})}\BibitemShut {NoStop}%
\bibitem [{\citenamefont {Mercurio}\ \emph {et~al.}(2023)\citenamefont
  {Mercurio}, \citenamefont {Andolina}, \citenamefont {Pellegrino},
  \citenamefont {Di~Stefano}, \citenamefont {Jarillo-Herrero}, \citenamefont
  {Felser}, \citenamefont {Koppens}, \citenamefont {Savasta},\ and\
  \citenamefont {Polini}}]{mercurio2023photon}%
  \BibitemOpen
  \bibfield  {author} {\bibinfo {author} {\bibfnamefont {A.}~\bibnamefont
  {Mercurio}}, \bibinfo {author} {\bibfnamefont {G.~M.}\ \bibnamefont
  {Andolina}}, \bibinfo {author} {\bibfnamefont {F.}~\bibnamefont
  {Pellegrino}}, \bibinfo {author} {\bibfnamefont {O.}~\bibnamefont
  {Di~Stefano}}, \bibinfo {author} {\bibfnamefont {P.}~\bibnamefont
  {Jarillo-Herrero}}, \bibinfo {author} {\bibfnamefont {C.}~\bibnamefont
  {Felser}}, \bibinfo {author} {\bibfnamefont {F.~H.}\ \bibnamefont {Koppens}},
  \bibinfo {author} {\bibfnamefont {S.}~\bibnamefont {Savasta}}, \ and\
  \bibinfo {author} {\bibfnamefont {M.}~\bibnamefont {Polini}},\ }\href@noop {}
  {\bibfield  {journal} {\bibinfo  {journal} {arXiv preprint arXiv:2302.09964}\
  } (\bibinfo {year} {2023})}\BibitemShut {NoStop}%
\bibitem [{\citenamefont {Vogl}\ \emph {et~al.}(2020)\citenamefont {Vogl},
  \citenamefont {Rodriguez-Vega},\ and\ \citenamefont
  {Fiete}}]{PhysRevB.101.241408}%
  \BibitemOpen
  \bibfield  {author} {\bibinfo {author} {\bibfnamefont {M.}~\bibnamefont
  {Vogl}}, \bibinfo {author} {\bibfnamefont {M.}~\bibnamefont
  {Rodriguez-Vega}}, \ and\ \bibinfo {author} {\bibfnamefont {G.~A.}\
  \bibnamefont {Fiete}},\ }\href {\doibase 10.1103/PhysRevB.101.241408}
  {\bibfield  {journal} {\bibinfo  {journal} {Phys. Rev. B}\ }\textbf {\bibinfo
  {volume} {101}},\ \bibinfo {pages} {241408} (\bibinfo {year}
  {2020})}\BibitemShut {NoStop}%
\bibitem [{\citenamefont {Topp}\ \emph {et~al.}(2019)\citenamefont {Topp},
  \citenamefont {Jotzu}, \citenamefont {McIver}, \citenamefont {Xian},
  \citenamefont {Rubio},\ and\ \citenamefont
  {Sentef}}]{PhysRevResearch.1.023031}%
  \BibitemOpen
  \bibfield  {author} {\bibinfo {author} {\bibfnamefont {G.~E.}\ \bibnamefont
  {Topp}}, \bibinfo {author} {\bibfnamefont {G.}~\bibnamefont {Jotzu}},
  \bibinfo {author} {\bibfnamefont {J.~W.}\ \bibnamefont {McIver}}, \bibinfo
  {author} {\bibfnamefont {L.}~\bibnamefont {Xian}}, \bibinfo {author}
  {\bibfnamefont {A.}~\bibnamefont {Rubio}}, \ and\ \bibinfo {author}
  {\bibfnamefont {M.~A.}\ \bibnamefont {Sentef}},\ }\href {\doibase
  10.1103/PhysRevResearch.1.023031} {\bibfield  {journal} {\bibinfo  {journal}
  {Phys. Rev. Res.}\ }\textbf {\bibinfo {volume} {1}},\ \bibinfo {pages}
  {023031} (\bibinfo {year} {2019})}\BibitemShut {NoStop}%
\bibitem [{\citenamefont {Wang}\ \emph {et~al.}(2013)\citenamefont {Wang},
  \citenamefont {Steinberg}, \citenamefont {Jarillo-Herrero},\ and\
  \citenamefont {Gedik}}]{doi:10.1126/science.1239834}%
  \BibitemOpen
  \bibfield  {author} {\bibinfo {author} {\bibfnamefont {Y.~H.}\ \bibnamefont
  {Wang}}, \bibinfo {author} {\bibfnamefont {H.}~\bibnamefont {Steinberg}},
  \bibinfo {author} {\bibfnamefont {P.}~\bibnamefont {Jarillo-Herrero}}, \ and\
  \bibinfo {author} {\bibfnamefont {N.}~\bibnamefont {Gedik}},\ }\href
  {\doibase 10.1126/science.1239834} {\bibfield  {journal} {\bibinfo  {journal}
  {Science}\ }\textbf {\bibinfo {volume} {342}},\ \bibinfo {pages} {453}
  (\bibinfo {year} {2013})}\BibitemShut {NoStop}%
\bibitem [{\citenamefont {Rokaj}\ \emph {et~al.}(2022)\citenamefont {Rokaj},
  \citenamefont {Ruggenthaler}, \citenamefont {Eich},\ and\ \citenamefont
  {Rubio}}]{rokaj2022free}%
  \BibitemOpen
  \bibfield  {author} {\bibinfo {author} {\bibfnamefont {V.}~\bibnamefont
  {Rokaj}}, \bibinfo {author} {\bibfnamefont {M.}~\bibnamefont {Ruggenthaler}},
  \bibinfo {author} {\bibfnamefont {F.~G.}\ \bibnamefont {Eich}}, \ and\
  \bibinfo {author} {\bibfnamefont {A.}~\bibnamefont {Rubio}},\ }\href@noop {}
  {\bibfield  {journal} {\bibinfo  {journal} {Physical Review Research}\
  }\textbf {\bibinfo {volume} {4}},\ \bibinfo {pages} {013012} (\bibinfo {year}
  {2022})}\BibitemShut {NoStop}%
\bibitem [{\citenamefont {Moddel}\ \emph {et~al.}(2021)\citenamefont {Moddel},
  \citenamefont {Weerakkody}, \citenamefont {Doroski},\ and\ \citenamefont
  {Bartusiak}}]{moddel2021casimir}%
  \BibitemOpen
  \bibfield  {author} {\bibinfo {author} {\bibfnamefont {G.}~\bibnamefont
  {Moddel}}, \bibinfo {author} {\bibfnamefont {A.}~\bibnamefont {Weerakkody}},
  \bibinfo {author} {\bibfnamefont {D.}~\bibnamefont {Doroski}}, \ and\
  \bibinfo {author} {\bibfnamefont {D.}~\bibnamefont {Bartusiak}},\ }\href@noop
  {} {\bibfield  {journal} {\bibinfo  {journal} {Physical Review Research}\
  }\textbf {\bibinfo {volume} {3}},\ \bibinfo {pages} {L022007} (\bibinfo
  {year} {2021})}\BibitemShut {NoStop}%
\bibitem [{\citenamefont {Sentef}\ \emph {et~al.}(2018)\citenamefont {Sentef},
  \citenamefont {Ruggenthaler},\ and\ \citenamefont
  {Rubio}}]{sentef2018cavity}%
  \BibitemOpen
  \bibfield  {author} {\bibinfo {author} {\bibfnamefont {M.~A.}\ \bibnamefont
  {Sentef}}, \bibinfo {author} {\bibfnamefont {M.}~\bibnamefont
  {Ruggenthaler}}, \ and\ \bibinfo {author} {\bibfnamefont {A.}~\bibnamefont
  {Rubio}},\ }\href@noop {} {\bibfield  {journal} {\bibinfo  {journal} {Science
  advances}\ }\textbf {\bibinfo {volume} {4}},\ \bibinfo {pages} {eaau6969}
  (\bibinfo {year} {2018})}\BibitemShut {NoStop}%
\bibitem [{\citenamefont {Schlawin}\ \emph {et~al.}(2019)\citenamefont
  {Schlawin}, \citenamefont {Cavalleri},\ and\ \citenamefont
  {Jaksch}}]{schlawin2019cavity}%
  \BibitemOpen
  \bibfield  {author} {\bibinfo {author} {\bibfnamefont {F.}~\bibnamefont
  {Schlawin}}, \bibinfo {author} {\bibfnamefont {A.}~\bibnamefont {Cavalleri}},
  \ and\ \bibinfo {author} {\bibfnamefont {D.}~\bibnamefont {Jaksch}},\
  }\href@noop {} {\bibfield  {journal} {\bibinfo  {journal} {Physical review
  letters}\ }\textbf {\bibinfo {volume} {122}},\ \bibinfo {pages} {133602}
  (\bibinfo {year} {2019})}\BibitemShut {NoStop}%
\bibitem [{\citenamefont {Curtis}\ \emph {et~al.}(2019)\citenamefont {Curtis},
  \citenamefont {Raines}, \citenamefont {Allocca}, \citenamefont {Hafezi},\
  and\ \citenamefont {Galitski}}]{curtis2019cavity}%
  \BibitemOpen
  \bibfield  {author} {\bibinfo {author} {\bibfnamefont {J.~B.}\ \bibnamefont
  {Curtis}}, \bibinfo {author} {\bibfnamefont {Z.~M.}\ \bibnamefont {Raines}},
  \bibinfo {author} {\bibfnamefont {A.~A.}\ \bibnamefont {Allocca}}, \bibinfo
  {author} {\bibfnamefont {M.}~\bibnamefont {Hafezi}}, \ and\ \bibinfo {author}
  {\bibfnamefont {V.~M.}\ \bibnamefont {Galitski}},\ }\href@noop {} {\bibfield
  {journal} {\bibinfo  {journal} {Physical review letters}\ }\textbf {\bibinfo
  {volume} {122}},\ \bibinfo {pages} {167002} (\bibinfo {year}
  {2019})}\BibitemShut {NoStop}%
\bibitem [{\citenamefont {Thomas}\ \emph {et~al.}(2019)\citenamefont {Thomas},
  \citenamefont {Devaux}, \citenamefont {Nagarajan}, \citenamefont {Chervy},
  \citenamefont {Seidel}, \citenamefont {Hagenm{\"u}ller}, \citenamefont
  {Sch{\"u}tz}, \citenamefont {Schachenmayer}, \citenamefont {Genet},
  \citenamefont {Pupillo} \emph {et~al.}}]{thomas2019exploring}%
  \BibitemOpen
  \bibfield  {author} {\bibinfo {author} {\bibfnamefont {A.}~\bibnamefont
  {Thomas}}, \bibinfo {author} {\bibfnamefont {E.}~\bibnamefont {Devaux}},
  \bibinfo {author} {\bibfnamefont {K.}~\bibnamefont {Nagarajan}}, \bibinfo
  {author} {\bibfnamefont {T.}~\bibnamefont {Chervy}}, \bibinfo {author}
  {\bibfnamefont {M.}~\bibnamefont {Seidel}}, \bibinfo {author} {\bibfnamefont
  {D.}~\bibnamefont {Hagenm{\"u}ller}}, \bibinfo {author} {\bibfnamefont
  {S.}~\bibnamefont {Sch{\"u}tz}}, \bibinfo {author} {\bibfnamefont
  {J.}~\bibnamefont {Schachenmayer}}, \bibinfo {author} {\bibfnamefont
  {C.}~\bibnamefont {Genet}}, \bibinfo {author} {\bibfnamefont
  {G.}~\bibnamefont {Pupillo}},  \emph {et~al.},\ }\href@noop {} {\bibfield
  {journal} {\bibinfo  {journal} {arXiv preprint arXiv:1911.01459}\ } (\bibinfo
  {year} {2019})}\BibitemShut {NoStop}%
\bibitem [{\citenamefont {Ciuti}(2021)}]{PhysRevB.104.155307}%
  \BibitemOpen
  \bibfield  {author} {\bibinfo {author} {\bibfnamefont {C.}~\bibnamefont
  {Ciuti}},\ }\href {\doibase 10.1103/PhysRevB.104.155307} {\bibfield
  {journal} {\bibinfo  {journal} {Phys. Rev. B}\ }\textbf {\bibinfo {volume}
  {104}},\ \bibinfo {pages} {155307} (\bibinfo {year} {2021})}\BibitemShut
  {NoStop}%
\bibitem [{\citenamefont {Espinosa-Ortega}\ \emph {et~al.}(2014)\citenamefont
  {Espinosa-Ortega}, \citenamefont {Kyriienko}, \citenamefont {Kibis},\ and\
  \citenamefont {Shelykh}}]{espinosa2014semiconductor}%
  \BibitemOpen
  \bibfield  {author} {\bibinfo {author} {\bibfnamefont {T.}~\bibnamefont
  {Espinosa-Ortega}}, \bibinfo {author} {\bibfnamefont {O.}~\bibnamefont
  {Kyriienko}}, \bibinfo {author} {\bibfnamefont {O.}~\bibnamefont {Kibis}}, \
  and\ \bibinfo {author} {\bibfnamefont {I.}~\bibnamefont {Shelykh}},\
  }\href@noop {} {\bibfield  {journal} {\bibinfo  {journal} {Physical Review
  A}\ }\textbf {\bibinfo {volume} {89}},\ \bibinfo {pages} {062115} (\bibinfo
  {year} {2014})}\BibitemShut {NoStop}%
\bibitem [{\citenamefont {Wang}\ \emph {et~al.}(2019)\citenamefont {Wang},
  \citenamefont {Ronca},\ and\ \citenamefont {Sentef}}]{PhysRevB.99.235156}%
  \BibitemOpen
  \bibfield  {author} {\bibinfo {author} {\bibfnamefont {X.}~\bibnamefont
  {Wang}}, \bibinfo {author} {\bibfnamefont {E.}~\bibnamefont {Ronca}}, \ and\
  \bibinfo {author} {\bibfnamefont {M.~A.}\ \bibnamefont {Sentef}},\ }\href
  {\doibase 10.1103/PhysRevB.99.235156} {\bibfield  {journal} {\bibinfo
  {journal} {Phys. Rev. B}\ }\textbf {\bibinfo {volume} {99}},\ \bibinfo
  {pages} {235156} (\bibinfo {year} {2019})}\BibitemShut {NoStop}%
\bibitem [{\citenamefont {Appugliese}\ \emph {et~al.}(2022)\citenamefont
  {Appugliese}, \citenamefont {Enkner}, \citenamefont {Paravicini-Bagliani},
  \citenamefont {Beck}, \citenamefont {Reichl}, \citenamefont {Wegscheider},
  \citenamefont {Scalari}, \citenamefont {Ciuti},\ and\ \citenamefont
  {Faist}}]{appugliese2022breakdown}%
  \BibitemOpen
  \bibfield  {author} {\bibinfo {author} {\bibfnamefont {F.}~\bibnamefont
  {Appugliese}}, \bibinfo {author} {\bibfnamefont {J.}~\bibnamefont {Enkner}},
  \bibinfo {author} {\bibfnamefont {G.~L.}\ \bibnamefont
  {Paravicini-Bagliani}}, \bibinfo {author} {\bibfnamefont {M.}~\bibnamefont
  {Beck}}, \bibinfo {author} {\bibfnamefont {C.}~\bibnamefont {Reichl}},
  \bibinfo {author} {\bibfnamefont {W.}~\bibnamefont {Wegscheider}}, \bibinfo
  {author} {\bibfnamefont {G.}~\bibnamefont {Scalari}}, \bibinfo {author}
  {\bibfnamefont {C.}~\bibnamefont {Ciuti}}, \ and\ \bibinfo {author}
  {\bibfnamefont {J.}~\bibnamefont {Faist}},\ }\href@noop {} {\bibfield
  {journal} {\bibinfo  {journal} {Science}\ }\textbf {\bibinfo {volume}
  {375}},\ \bibinfo {pages} {1030} (\bibinfo {year} {2022})}\BibitemShut
  {NoStop}%
\bibitem [{\citenamefont {Galego}\ \emph {et~al.}(2019)\citenamefont {Galego},
  \citenamefont {Climent}, \citenamefont {Garcia-Vidal},\ and\ \citenamefont
  {Feist}}]{galego2019cavity}%
  \BibitemOpen
  \bibfield  {author} {\bibinfo {author} {\bibfnamefont {J.}~\bibnamefont
  {Galego}}, \bibinfo {author} {\bibfnamefont {C.}~\bibnamefont {Climent}},
  \bibinfo {author} {\bibfnamefont {F.~J.}\ \bibnamefont {Garcia-Vidal}}, \
  and\ \bibinfo {author} {\bibfnamefont {J.}~\bibnamefont {Feist}},\
  }\href@noop {} {\bibfield  {journal} {\bibinfo  {journal} {Physical Review
  X}\ }\textbf {\bibinfo {volume} {9}},\ \bibinfo {pages} {021057} (\bibinfo
  {year} {2019})}\BibitemShut {NoStop}%
\bibitem [{\citenamefont {Galego}\ \emph {et~al.}(2015)\citenamefont {Galego},
  \citenamefont {Garcia-Vidal},\ and\ \citenamefont
  {Feist}}]{galego2015cavity}%
  \BibitemOpen
  \bibfield  {author} {\bibinfo {author} {\bibfnamefont {J.}~\bibnamefont
  {Galego}}, \bibinfo {author} {\bibfnamefont {F.~J.}\ \bibnamefont
  {Garcia-Vidal}}, \ and\ \bibinfo {author} {\bibfnamefont {J.}~\bibnamefont
  {Feist}},\ }\href@noop {} {\bibfield  {journal} {\bibinfo  {journal}
  {Physical Review X}\ }\textbf {\bibinfo {volume} {5}},\ \bibinfo {pages}
  {041022} (\bibinfo {year} {2015})}\BibitemShut {NoStop}%
\bibitem [{\citenamefont {Long}\ \emph {et~al.}(2022)\citenamefont {Long},
  \citenamefont {Pantale{\'o}n}, \citenamefont {Zhan}, \citenamefont {Guinea},
  \citenamefont {Silva-Guill{\'e}n},\ and\ \citenamefont {Yuan}}]{Long2022}%
  \BibitemOpen
  \bibfield  {author} {\bibinfo {author} {\bibfnamefont {M.}~\bibnamefont
  {Long}}, \bibinfo {author} {\bibfnamefont {P.~A.}\ \bibnamefont
  {Pantale{\'o}n}}, \bibinfo {author} {\bibfnamefont {Z.}~\bibnamefont {Zhan}},
  \bibinfo {author} {\bibfnamefont {F.}~\bibnamefont {Guinea}}, \bibinfo
  {author} {\bibfnamefont {J.~{\'A}.}\ \bibnamefont {Silva-Guill{\'e}n}}, \
  and\ \bibinfo {author} {\bibfnamefont {S.}~\bibnamefont {Yuan}},\ }\href
  {\doibase 10.1038/s41524-022-00763-1} {\bibfield  {journal} {\bibinfo
  {journal} {npj Computational Materials}\ }\textbf {\bibinfo {volume} {8}},\
  \bibinfo {pages} {73} (\bibinfo {year} {2022})}\BibitemShut {NoStop}%
\bibitem [{\citenamefont {Butcher}\ \emph {et~al.}(2012)\citenamefont
  {Butcher}, \citenamefont {Buhmann},\ and\ \citenamefont
  {Scheel}}]{butcher2012casimir}%
  \BibitemOpen
  \bibfield  {author} {\bibinfo {author} {\bibfnamefont {D.~T.}\ \bibnamefont
  {Butcher}}, \bibinfo {author} {\bibfnamefont {S.~Y.}\ \bibnamefont
  {Buhmann}}, \ and\ \bibinfo {author} {\bibfnamefont {S.}~\bibnamefont
  {Scheel}},\ }\href@noop {} {\bibfield  {journal} {\bibinfo  {journal} {New
  Journal of Physics}\ }\textbf {\bibinfo {volume} {14}},\ \bibinfo {pages}
  {113013} (\bibinfo {year} {2012})}\BibitemShut {NoStop}%
\bibitem [{\citenamefont {Jiang}\ and\ \citenamefont
  {Wilczek}(2019{\natexlab{a}})}]{jiang2019axial}%
  \BibitemOpen
  \bibfield  {author} {\bibinfo {author} {\bibfnamefont {Q.-D.}\ \bibnamefont
  {Jiang}}\ and\ \bibinfo {author} {\bibfnamefont {F.}~\bibnamefont
  {Wilczek}},\ }\href@noop {} {\bibfield  {journal} {\bibinfo  {journal}
  {Physical Review B}\ }\textbf {\bibinfo {volume} {99}},\ \bibinfo {pages}
  {165402} (\bibinfo {year} {2019}{\natexlab{a}})}\BibitemShut {NoStop}%
\bibitem [{\citenamefont {Jiang}(2023)}]{jiang2023angular}%
  \BibitemOpen
  \bibfield  {author} {\bibinfo {author} {\bibfnamefont {Q.-D.}\ \bibnamefont
  {Jiang}},\ }\href@noop {} {\bibfield  {journal} {\bibinfo  {journal} {arXiv
  preprint arXiv:2307.14964}\ } (\bibinfo {year} {2023})}\BibitemShut {NoStop}%
\bibitem [{\citenamefont {Ke}\ \emph {et~al.}(2023)\citenamefont {Ke},
  \citenamefont {Song},\ and\ \citenamefont {Jiang}}]{PhysRevLett.131.223601}%
  \BibitemOpen
  \bibfield  {author} {\bibinfo {author} {\bibfnamefont {Y.}~\bibnamefont
  {Ke}}, \bibinfo {author} {\bibfnamefont {Z.}~\bibnamefont {Song}}, \ and\
  \bibinfo {author} {\bibfnamefont {Q.-D.}\ \bibnamefont {Jiang}},\ }\href
  {\doibase 10.1103/PhysRevLett.131.223601} {\bibfield  {journal} {\bibinfo
  {journal} {Phys. Rev. Lett.}\ }\textbf {\bibinfo {volume} {131}},\ \bibinfo
  {pages} {223601} (\bibinfo {year} {2023})}\BibitemShut {NoStop}%
\bibitem [{\citenamefont {Riso}\ \emph {et~al.}(2023)\citenamefont {Riso},
  \citenamefont {Grazioli}, \citenamefont {Ronca}, \citenamefont {Giovannini},\
  and\ \citenamefont {Koch}}]{riso2023strong}%
  \BibitemOpen
  \bibfield  {author} {\bibinfo {author} {\bibfnamefont {R.~R.}\ \bibnamefont
  {Riso}}, \bibinfo {author} {\bibfnamefont {L.}~\bibnamefont {Grazioli}},
  \bibinfo {author} {\bibfnamefont {E.}~\bibnamefont {Ronca}}, \bibinfo
  {author} {\bibfnamefont {T.}~\bibnamefont {Giovannini}}, \ and\ \bibinfo
  {author} {\bibfnamefont {H.}~\bibnamefont {Koch}},\ }\href@noop {} {\bibfield
   {journal} {\bibinfo  {journal} {Physical Review X}\ }\textbf {\bibinfo
  {volume} {13}},\ \bibinfo {pages} {031002} (\bibinfo {year}
  {2023})}\BibitemShut {NoStop}%
\bibitem [{\citenamefont {Vu}\ \emph {et~al.}(2022)\citenamefont {Vu},
  \citenamefont {McLeod}, \citenamefont {Hanson},\ and\ \citenamefont
  {DePrince~III}}]{vu2023enhanced}%
  \BibitemOpen
  \bibfield  {author} {\bibinfo {author} {\bibfnamefont {N.}~\bibnamefont
  {Vu}}, \bibinfo {author} {\bibfnamefont {G.~M.}\ \bibnamefont {McLeod}},
  \bibinfo {author} {\bibfnamefont {K.}~\bibnamefont {Hanson}}, \ and\ \bibinfo
  {author} {\bibfnamefont {A.~E.}\ \bibnamefont {DePrince~III}},\ }\href@noop
  {} {\bibfield  {journal} {\bibinfo  {journal} {The Journal of Physical
  Chemistry A}\ }\textbf {\bibinfo {volume} {126}},\ \bibinfo {pages} {9303}
  (\bibinfo {year} {2022})}\BibitemShut {NoStop}%
\bibitem [{\citenamefont {Jiang}\ and\ \citenamefont
  {Wilczek}(2019{\natexlab{b}})}]{jiang2019quantum}%
  \BibitemOpen
  \bibfield  {author} {\bibinfo {author} {\bibfnamefont {Q.-D.}\ \bibnamefont
  {Jiang}}\ and\ \bibinfo {author} {\bibfnamefont {F.}~\bibnamefont
  {Wilczek}},\ }\href@noop {} {\bibfield  {journal} {\bibinfo  {journal}
  {Physical Review B}\ }\textbf {\bibinfo {volume} {99}},\ \bibinfo {pages}
  {201104} (\bibinfo {year} {2019}{\natexlab{b}})}\BibitemShut {NoStop}%
\bibitem [{\citenamefont {Catarina}\ \emph {et~al.}(2019)\citenamefont
  {Catarina}, \citenamefont {Amorim}, \citenamefont {Castro}, \citenamefont
  {Lopes}, \citenamefont {Lopes},\ and\ \citenamefont {Peres}}]{handbook}%
  \BibitemOpen
  \bibfield  {author} {\bibinfo {author} {\bibfnamefont {G.}~\bibnamefont
  {Catarina}}, \bibinfo {author} {\bibfnamefont {B.}~\bibnamefont {Amorim}},
  \bibinfo {author} {\bibfnamefont {E.~V.}\ \bibnamefont {Castro}}, \bibinfo
  {author} {\bibfnamefont {J.~M. V.~P.}\ \bibnamefont {Lopes}}, \bibinfo
  {author} {\bibfnamefont {J.~M. V.~P.}\ \bibnamefont {Lopes}}, \ and\ \bibinfo
  {author} {\bibfnamefont {N.}~\bibnamefont {Peres}},\ }\enquote {\bibinfo
  {title} {Twisted bilayer graphene: Low-energy physics, electronic and optical
  properties},}\ in\ \href {\doibase
  https://doi.org/10.1002/9781119468455.ch44} {\emph {\bibinfo {booktitle}
  {Handbook of Graphene Set}}}\ (\bibinfo  {publisher} {John Wiley \& Sons,
  Ltd},\ \bibinfo {year} {2019})\ Chap.~\bibinfo {chapter} {6}, pp.\ \bibinfo
  {pages} {177--231}\BibitemShut {NoStop}%
\bibitem [{\citenamefont {Uchida}\ \emph {et~al.}(2014)\citenamefont {Uchida},
  \citenamefont {Furuya}, \citenamefont {Iwata},\ and\ \citenamefont
  {Oshiyama}}]{PhysRevB.90.155451}%
  \BibitemOpen
  \bibfield  {author} {\bibinfo {author} {\bibfnamefont {K.}~\bibnamefont
  {Uchida}}, \bibinfo {author} {\bibfnamefont {S.}~\bibnamefont {Furuya}},
  \bibinfo {author} {\bibfnamefont {J.-I.}\ \bibnamefont {Iwata}}, \ and\
  \bibinfo {author} {\bibfnamefont {A.}~\bibnamefont {Oshiyama}},\ }\href
  {\doibase 10.1103/PhysRevB.90.155451} {\bibfield  {journal} {\bibinfo
  {journal} {Phys. Rev. B}\ }\textbf {\bibinfo {volume} {90}},\ \bibinfo
  {pages} {155451} (\bibinfo {year} {2014})}\BibitemShut {NoStop}%
\bibitem [{\citenamefont {Lucignano}\ \emph {et~al.}(2019)\citenamefont
  {Lucignano}, \citenamefont {Alf\`e}, \citenamefont {Cataudella},
  \citenamefont {Ninno},\ and\ \citenamefont {Cantele}}]{PhysRevB.99.195419}%
  \BibitemOpen
  \bibfield  {author} {\bibinfo {author} {\bibfnamefont {P.}~\bibnamefont
  {Lucignano}}, \bibinfo {author} {\bibfnamefont {D.}~\bibnamefont {Alf\`e}},
  \bibinfo {author} {\bibfnamefont {V.}~\bibnamefont {Cataudella}}, \bibinfo
  {author} {\bibfnamefont {D.}~\bibnamefont {Ninno}}, \ and\ \bibinfo {author}
  {\bibfnamefont {G.}~\bibnamefont {Cantele}},\ }\href {\doibase
  10.1103/PhysRevB.99.195419} {\bibfield  {journal} {\bibinfo  {journal} {Phys.
  Rev. B}\ }\textbf {\bibinfo {volume} {99}},\ \bibinfo {pages} {195419}
  (\bibinfo {year} {2019})}\BibitemShut {NoStop}%
\bibitem [{\citenamefont {Bistritzer}\ and\ \citenamefont
  {MacDonald}(2011)}]{doi:10.1073/pnas.1108174108}%
  \BibitemOpen
  \bibfield  {author} {\bibinfo {author} {\bibfnamefont {R.}~\bibnamefont
  {Bistritzer}}\ and\ \bibinfo {author} {\bibfnamefont {A.~H.}\ \bibnamefont
  {MacDonald}},\ }\href {\doibase 10.1073/pnas.1108174108} {\bibfield
  {journal} {\bibinfo  {journal} {Proceedings of the National Academy of
  Sciences}\ }\textbf {\bibinfo {volume} {108}},\ \bibinfo {pages} {12233}
  (\bibinfo {year} {2011})}\BibitemShut {NoStop}%
\bibitem [{\citenamefont {Su\'arez~Morell}\ \emph {et~al.}(2010)\citenamefont
  {Su\'arez~Morell}, \citenamefont {Correa}, \citenamefont {Vargas},
  \citenamefont {Pacheco},\ and\ \citenamefont
  {Barticevic}}]{PhysRevB.82.121407}%
  \BibitemOpen
  \bibfield  {author} {\bibinfo {author} {\bibfnamefont {E.}~\bibnamefont
  {Su\'arez~Morell}}, \bibinfo {author} {\bibfnamefont {J.~D.}\ \bibnamefont
  {Correa}}, \bibinfo {author} {\bibfnamefont {P.}~\bibnamefont {Vargas}},
  \bibinfo {author} {\bibfnamefont {M.}~\bibnamefont {Pacheco}}, \ and\
  \bibinfo {author} {\bibfnamefont {Z.}~\bibnamefont {Barticevic}},\ }\href
  {\doibase 10.1103/PhysRevB.82.121407} {\bibfield  {journal} {\bibinfo
  {journal} {Phys. Rev. B}\ }\textbf {\bibinfo {volume} {82}},\ \bibinfo
  {pages} {121407} (\bibinfo {year} {2010})}\BibitemShut {NoStop}%
\bibitem [{\citenamefont {Andrei}\ and\ \citenamefont
  {MacDonald}(2020)}]{Andrei2020}%
  \BibitemOpen
  \bibfield  {author} {\bibinfo {author} {\bibfnamefont {E.~Y.}\ \bibnamefont
  {Andrei}}\ and\ \bibinfo {author} {\bibfnamefont {A.~H.}\ \bibnamefont
  {MacDonald}},\ }\href {\doibase 10.1038/s41563-020-00840-0} {\bibfield
  {journal} {\bibinfo  {journal} {Nature Materials}\ }\textbf {\bibinfo
  {volume} {19}},\ \bibinfo {pages} {1265} (\bibinfo {year}
  {2020})}\BibitemShut {NoStop}%
\bibitem [{\citenamefont {Song}\ \emph {et~al.}(2005)\citenamefont {Song},
  \citenamefont {Noda}, \citenamefont {Asano},\ and\ \citenamefont
  {Akahane}}]{Song2005}%
  \BibitemOpen
  \bibfield  {author} {\bibinfo {author} {\bibfnamefont {B.-S.}\ \bibnamefont
  {Song}}, \bibinfo {author} {\bibfnamefont {S.}~\bibnamefont {Noda}}, \bibinfo
  {author} {\bibfnamefont {T.}~\bibnamefont {Asano}}, \ and\ \bibinfo {author}
  {\bibfnamefont {Y.}~\bibnamefont {Akahane}},\ }\href {\doibase
  10.1038/nmat1320} {\bibfield  {journal} {\bibinfo  {journal} {Nature
  Materials}\ }\textbf {\bibinfo {volume} {4}},\ \bibinfo {pages} {207}
  (\bibinfo {year} {2005})}\BibitemShut {NoStop}%
\bibitem [{\citenamefont {Xiao}\ \emph {et~al.}(2010)\citenamefont {Xiao},
  \citenamefont {Chang},\ and\ \citenamefont {Niu}}]{RevModPhys.82.1959}%
  \BibitemOpen
  \bibfield  {author} {\bibinfo {author} {\bibfnamefont {D.}~\bibnamefont
  {Xiao}}, \bibinfo {author} {\bibfnamefont {M.-C.}\ \bibnamefont {Chang}}, \
  and\ \bibinfo {author} {\bibfnamefont {Q.}~\bibnamefont {Niu}},\ }\href
  {\doibase 10.1103/RevModPhys.82.1959} {\bibfield  {journal} {\bibinfo
  {journal} {Rev. Mod. Phys.}\ }\textbf {\bibinfo {volume} {82}},\ \bibinfo
  {pages} {1959} (\bibinfo {year} {2010})}\BibitemShut {NoStop}%
\bibitem [{\citenamefont {Sodemann}\ and\ \citenamefont
  {Fu}(2015)}]{PhysRevLett.115.216806}%
  \BibitemOpen
  \bibfield  {author} {\bibinfo {author} {\bibfnamefont {I.}~\bibnamefont
  {Sodemann}}\ and\ \bibinfo {author} {\bibfnamefont {L.}~\bibnamefont {Fu}},\
  }\href {\doibase 10.1103/PhysRevLett.115.216806} {\bibfield  {journal}
  {\bibinfo  {journal} {Phys. Rev. Lett.}\ }\textbf {\bibinfo {volume} {115}},\
  \bibinfo {pages} {216806} (\bibinfo {year} {2015})}\BibitemShut {NoStop}%
\bibitem [{\citenamefont {Maissen}\ \emph {et~al.}(2014)\citenamefont
  {Maissen}, \citenamefont {Scalari}, \citenamefont {Valmorra}, \citenamefont
  {Beck}, \citenamefont {Faist}, \citenamefont {Cibella}, \citenamefont
  {Leoni}, \citenamefont {Reichl}, \citenamefont {Charpentier},\ and\
  \citenamefont {Wegscheider}}]{PhysRevB.90.205309}%
  \BibitemOpen
  \bibfield  {author} {\bibinfo {author} {\bibfnamefont {C.}~\bibnamefont
  {Maissen}}, \bibinfo {author} {\bibfnamefont {G.}~\bibnamefont {Scalari}},
  \bibinfo {author} {\bibfnamefont {F.}~\bibnamefont {Valmorra}}, \bibinfo
  {author} {\bibfnamefont {M.}~\bibnamefont {Beck}}, \bibinfo {author}
  {\bibfnamefont {J.}~\bibnamefont {Faist}}, \bibinfo {author} {\bibfnamefont
  {S.}~\bibnamefont {Cibella}}, \bibinfo {author} {\bibfnamefont
  {R.}~\bibnamefont {Leoni}}, \bibinfo {author} {\bibfnamefont
  {C.}~\bibnamefont {Reichl}}, \bibinfo {author} {\bibfnamefont
  {C.}~\bibnamefont {Charpentier}}, \ and\ \bibinfo {author} {\bibfnamefont
  {W.}~\bibnamefont {Wegscheider}},\ }\href {\doibase
  10.1103/PhysRevB.90.205309} {\bibfield  {journal} {\bibinfo  {journal} {Phys.
  Rev. B}\ }\textbf {\bibinfo {volume} {90}},\ \bibinfo {pages} {205309}
  (\bibinfo {year} {2014})}\BibitemShut {NoStop}%
\bibitem [{\citenamefont {Frisk~Kockum}\ \emph {et~al.}(2019)\citenamefont
  {Frisk~Kockum}, \citenamefont {Miranowicz}, \citenamefont {De~Liberato},
  \citenamefont {Savasta},\ and\ \citenamefont {Nori}}]{FriskKockum2019}%
  \BibitemOpen
  \bibfield  {author} {\bibinfo {author} {\bibfnamefont {A.}~\bibnamefont
  {Frisk~Kockum}}, \bibinfo {author} {\bibfnamefont {A.}~\bibnamefont
  {Miranowicz}}, \bibinfo {author} {\bibfnamefont {S.}~\bibnamefont
  {De~Liberato}}, \bibinfo {author} {\bibfnamefont {S.}~\bibnamefont
  {Savasta}}, \ and\ \bibinfo {author} {\bibfnamefont {F.}~\bibnamefont
  {Nori}},\ }\href {\doibase 10.1038/s42254-018-0006-2} {\bibfield  {journal}
  {\bibinfo  {journal} {Nature Reviews Physics}\ }\textbf {\bibinfo {volume}
  {1}},\ \bibinfo {pages} {19} (\bibinfo {year} {2019})}\BibitemShut {NoStop}%
\bibitem [{\citenamefont {Scalari}\ \emph {et~al.}(2013)\citenamefont
  {Scalari}, \citenamefont {Maissen}, \citenamefont {Hagenm{\"u}ller},
  \citenamefont {De~Liberato}, \citenamefont {Ciuti}, \citenamefont {Reichl},
  \citenamefont {Wegscheider}, \citenamefont {Schuh}, \citenamefont {Beck},\
  and\ \citenamefont {Faist}}]{scalari2013ultrastrong}%
  \BibitemOpen
  \bibfield  {author} {\bibinfo {author} {\bibfnamefont {G.}~\bibnamefont
  {Scalari}}, \bibinfo {author} {\bibfnamefont {C.}~\bibnamefont {Maissen}},
  \bibinfo {author} {\bibfnamefont {D.}~\bibnamefont {Hagenm{\"u}ller}},
  \bibinfo {author} {\bibfnamefont {S.}~\bibnamefont {De~Liberato}}, \bibinfo
  {author} {\bibfnamefont {C.}~\bibnamefont {Ciuti}}, \bibinfo {author}
  {\bibfnamefont {C.}~\bibnamefont {Reichl}}, \bibinfo {author} {\bibfnamefont
  {W.}~\bibnamefont {Wegscheider}}, \bibinfo {author} {\bibfnamefont
  {D.}~\bibnamefont {Schuh}}, \bibinfo {author} {\bibfnamefont
  {M.}~\bibnamefont {Beck}}, \ and\ \bibinfo {author} {\bibfnamefont
  {J.}~\bibnamefont {Faist}},\ }\href@noop {} {\bibfield  {journal} {\bibinfo
  {journal} {Journal of Applied Physics}\ }\textbf {\bibinfo {volume} {113}}
  (\bibinfo {year} {2013})}\BibitemShut {NoStop}%
\bibitem [{\citenamefont {Mauro}\ \emph {et~al.}(2023)\citenamefont {Mauro},
  \citenamefont {Fregoni}, \citenamefont {Feist},\ and\ \citenamefont
  {Avriller}}]{PhysRevA.107.L021501}%
  \BibitemOpen
  \bibfield  {author} {\bibinfo {author} {\bibfnamefont {L.}~\bibnamefont
  {Mauro}}, \bibinfo {author} {\bibfnamefont {J.}~\bibnamefont {Fregoni}},
  \bibinfo {author} {\bibfnamefont {J.}~\bibnamefont {Feist}}, \ and\ \bibinfo
  {author} {\bibfnamefont {R.}~\bibnamefont {Avriller}},\ }\href {\doibase
  10.1103/PhysRevA.107.L021501} {\bibfield  {journal} {\bibinfo  {journal}
  {Phys. Rev. A}\ }\textbf {\bibinfo {volume} {107}},\ \bibinfo {pages}
  {L021501} (\bibinfo {year} {2023})}\BibitemShut {NoStop}%
\bibitem [{\citenamefont {Jarc}\ \emph {et~al.}(2023)\citenamefont {Jarc},
  \citenamefont {Mathengattil}, \citenamefont {Montanaro}, \citenamefont
  {Giusti}, \citenamefont {Rigoni}, \citenamefont {Sergo}, \citenamefont
  {Fassioli}, \citenamefont {Winnerl}, \citenamefont {Dal~Zilio}, \citenamefont
  {Mihailovic} \emph {et~al.}}]{jarc2023cavity}%
  \BibitemOpen
  \bibfield  {author} {\bibinfo {author} {\bibfnamefont {G.}~\bibnamefont
  {Jarc}}, \bibinfo {author} {\bibfnamefont {S.~Y.}\ \bibnamefont
  {Mathengattil}}, \bibinfo {author} {\bibfnamefont {A.}~\bibnamefont
  {Montanaro}}, \bibinfo {author} {\bibfnamefont {F.}~\bibnamefont {Giusti}},
  \bibinfo {author} {\bibfnamefont {E.~M.}\ \bibnamefont {Rigoni}}, \bibinfo
  {author} {\bibfnamefont {R.}~\bibnamefont {Sergo}}, \bibinfo {author}
  {\bibfnamefont {F.}~\bibnamefont {Fassioli}}, \bibinfo {author}
  {\bibfnamefont {S.}~\bibnamefont {Winnerl}}, \bibinfo {author} {\bibfnamefont
  {S.}~\bibnamefont {Dal~Zilio}}, \bibinfo {author} {\bibfnamefont
  {D.}~\bibnamefont {Mihailovic}},  \emph {et~al.},\ }\href@noop {} {\bibfield
  {journal} {\bibinfo  {journal} {Nature}\ }\textbf {\bibinfo {volume} {622}},\
  \bibinfo {pages} {487} (\bibinfo {year} {2023})}\BibitemShut {NoStop}%
\bibitem [{\citenamefont {Topp}\ \emph {et~al.}(2021)\citenamefont {Topp},
  \citenamefont {Eckhardt}, \citenamefont {Kennes}, \citenamefont {Sentef},\
  and\ \citenamefont {T\"orm\"a}}]{PhysRevB.104.064306}%
  \BibitemOpen
  \bibfield  {author} {\bibinfo {author} {\bibfnamefont {G.~E.}\ \bibnamefont
  {Topp}}, \bibinfo {author} {\bibfnamefont {C.~J.}\ \bibnamefont {Eckhardt}},
  \bibinfo {author} {\bibfnamefont {D.~M.}\ \bibnamefont {Kennes}}, \bibinfo
  {author} {\bibfnamefont {M.~A.}\ \bibnamefont {Sentef}}, \ and\ \bibinfo
  {author} {\bibfnamefont {P.}~\bibnamefont {T\"orm\"a}},\ }\href {\doibase
  10.1103/PhysRevB.104.064306} {\bibfield  {journal} {\bibinfo  {journal}
  {Phys. Rev. B}\ }\textbf {\bibinfo {volume} {104}},\ \bibinfo {pages}
  {064306} (\bibinfo {year} {2021})}\BibitemShut {NoStop}%
\bibitem [{\citenamefont {Tai}\ and\ \citenamefont
  {Claassen}(2023)}]{tai2023quantum}%
  \BibitemOpen
  \bibfield  {author} {\bibinfo {author} {\bibfnamefont {W.~T.}\ \bibnamefont
  {Tai}}\ and\ \bibinfo {author} {\bibfnamefont {M.}~\bibnamefont {Claassen}},\
  }\href@noop {} {\bibfield  {journal} {\bibinfo  {journal} {arXiv preprint
  arXiv:2303.01597}\ } (\bibinfo {year} {2023})}\BibitemShut {NoStop}%
\end{thebibliography}%

\newpage
\onecolumngrid
\appendix 
\clearpage
\renewcommand\thefigure{S\arabic{figure}}    
\setcounter{figure}{0} 
\renewcommand{\theequation}{S\arabic{equation}}
\setcounter{equation}{0}
\renewcommand{\thesubsection}{SM\arabic{subsection}}
\section*{\Large Supplementary Information}
In this Supplementary Information (SI), we provide further details about the numerical and analytical computations presented in the main text. Moreover, we show further analysis to corroborate our findings.

{\hypersetup{linkcolor=blue}
  \tableofcontents}

\subsection{Theoretical setup and numerical methods}

The Hamiltonian of monolayer graphene can be obtained by considering the hopping of electrons from one carbon atom to its three nearest neighboring atoms. The electronic band structure shows a characteristic linear dispersion relation around the Fermi energy. The low energy Hamiltonian reads \cite{handbook}
\begin{equation}
    H_{1}(\mathbf{q})=v_F \left(\begin{matrix}
    0 & q_x-i q_y \\ q_x+i q_y & 0
    \end{matrix}\right),
\end{equation}
where $v_F$ is the Fermi velocity which is around $v_F\approx 5.944 eV \cdot \angstrom$.

On the other hand, twisted bilayer graphene (TBG) is more complicated because the unit cell of the moir\'e pattern is much larger than that of graphene. Hence, the moir\'e Brillouin zone is quite small in reciprocal space and one should consider many hopping channels between a group of moir\'e Brillouin zones in the two layers. More details on this point can be found in \cite{handbook}. In this manuscript, we use the following effective Hamiltonian for TBG,
\begin{equation}
    H_{TBG}(\mathbf{q})=\left(\begin{matrix}
    H_1 (\mathbf{q}) & T_{\mathbf{q}_b} & T_{\mathbf{q}_{tr}} & T_{\mathbf{q}_{tl}} & \cdots\\ T_{\mathbf{q}_b}^\dagger &  H_2 (\mathbf{q}-\mathbf{q}_b) & 0 & 0 & \cdots\\ T_{\mathbf{q}_{tr}}^\dagger & 0 & H_2 (\mathbf{q}-\mathbf{q}_{tr}) & 0  & \cdots\\ T_{\mathbf{q}_{tl}}^\dagger & 0 & 0 & H_2 (\mathbf{q}-\mathbf{q}_{tl}) \\ \vdots & \vdots & \vdots & \vdots & \ddots
    \end{matrix}\right).\label{h}
\end{equation}
Here, $H_{1,2}(\mathbf{q})$ indicate the Hamiltonian of the top/bottom layers. Moreover,
\begin{equation}
    \mathbf{q}_b=\dfrac{1}{3}(\mathbf{b}_1^m-\mathbf{b}_2^m),\qquad \mathbf{q}_{tr}=\dfrac{1}{3}(\mathbf{b}_1^m+2\mathbf{b}_2^m), \qquad \mathbf{q}_{tl}=\dfrac{1}{3}(-2\mathbf{b}_1^m-\mathbf{b}_2^m),
\end{equation}
where $\mathbf{b}_1^m$ and $\mathbf{b}_2^m$ are the moir\`e reciprocal vectors containing the information about the twisting angle $\theta$. The hopping matrix elements are given by,
\begin{equation}
    T_{\mathbf{q}_{b}}= t\left(\begin{matrix}
    u & u' \\ u' & u
    \end{matrix}\right),\quad T_{\mathbf{q}_{tr}}= t\left(\begin{matrix}
    u e^{i \phi} & u' \\ u' e^{-i \phi} & u e^{i \phi}
    \end{matrix}\right),\quad T_{\mathbf{q}_{tl}}= t\left(\begin{matrix}
    u e^{-i \phi} & u' \\ u' e^{i \phi} & u e^{-i \phi}
    \end{matrix}\right),
\end{equation}
with $u=0.817$, $u'=1$, $\phi=2\pi/3$ and $t=0.11$ is the hopping parameter. This Hamiltonian takes into account that the AA and AB stacking regions have a different interlayer distance due to presence of corrugation \cite{PhysRevB.99.195419,PhysRevB.90.155451} and therefore different coupling strength \cite{PhysRevX.8.031087}. 

\begin{figure}
    \centering
    \includegraphics[width=\linewidth]{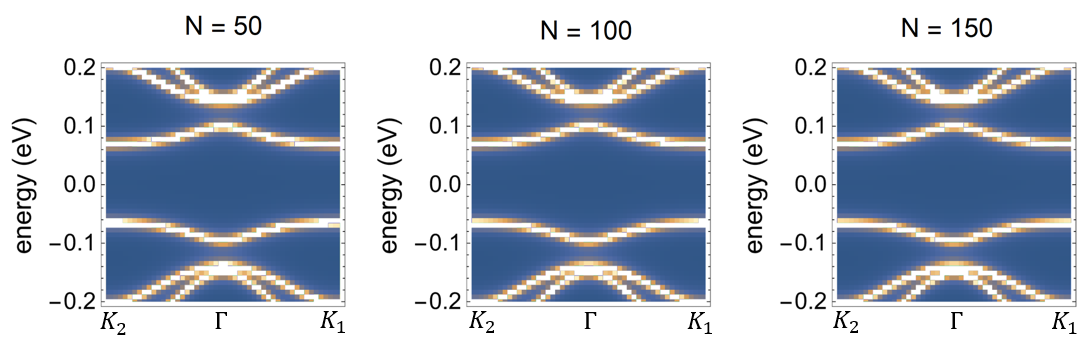}
    \caption{The results for the self-energy Eq.\eqref{self2} using different values of the cutoff $N$. From left to right, $N=50$, $N=100$, $N=150$ while $\theta=1.5^{\circ}$, $g=2$. This figure emphasizes the stability of our numerical routine and its convergence as a function of the cutoff $N$.}
    \label{fig:cutoff}
\end{figure}

As outlined in the main text, the self-energy is given by an infinite sum over the Matsubara frequencies as
\begin{equation}
    \Sigma_0 (\epsilon,\mathbf{q}) =-\dfrac{g^2}{\beta}\sum_{m=1}^\infty G_0 (\epsilon+i\omega_m,\mathbf{q}) D_0 (\omega_m),\label{sum}
\end{equation}
where $\omega_m=2 \pi m k_B T$ is the Matsubara frequency and $g$ the coupling parameter between photon and electron. Nevertheless, from a computational perspective, the Matsubara frequency summation in Eq.\eqref{sum} is performed by introducing a finite UV cutoff $N$
 \begin{equation}
    \Sigma_0 (\epsilon,\mathbf{q}) =-\dfrac{g^2}{\beta}\sum_{m=1}^N G_0 (\epsilon+i\omega_m,\mathbf{q}) D_0 (\omega_m).\label{self2}
\end{equation}
The computations shown in the main text are for $N=100$. To make sure that the sum converged for the value of the cutoff used in the main text, we have computed the same spectral function with different cutoffs as shown in Fig.\ref{fig:cutoff}. The results show only little difference upon changing the cutoff from $N=50$ to $N=150$ confirming that the numerical method is reliable.

\begin{figure}[htb]
    \centering
    \includegraphics[width=0.3\linewidth]{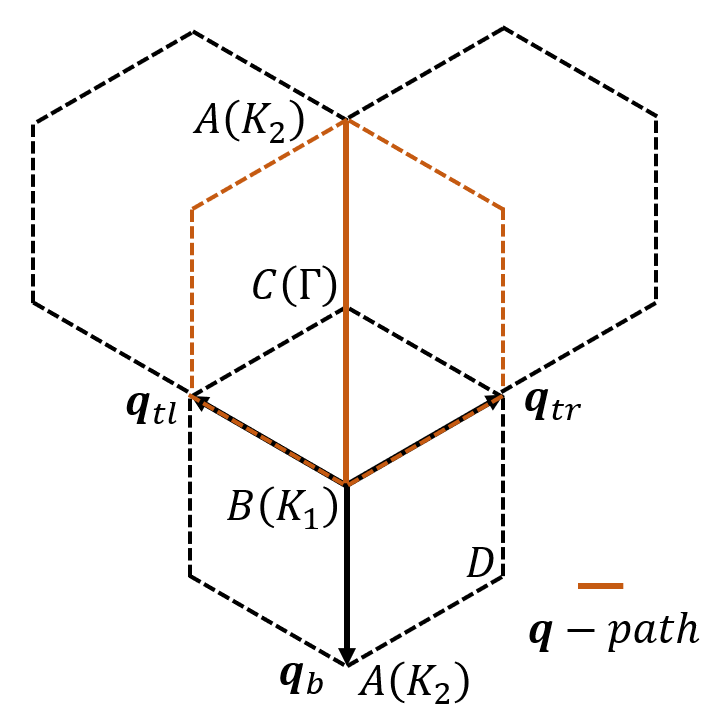}
    \caption{The moir\'e Brillouin zone with the high-symmetry points and the $q$-path indicated.}
    \label{BZ}
\end{figure}

For completeness, in Fig.\ref{BZ} we show the moir\'e Brillouin zone in reciprocal space where the high-symmetry points and $q$-path used in our computations are emphasized.

\subsection{The simpler case of graphene}
In this section, as a warp-up exercise, we revisit the effects of a chiral optical cavity on monolayer graphene (see \cite{PhysRevB.99.235156} for the original results). As for the case of twisted-bilayer graphene (TBG) presented in the main text, a simple theoretical model based on the breaking of time reversal symmetry will play an important role in understanding the outcomes of our computations.

\begin{figure}[hb]
    \centering
    \includegraphics[width=0.6\linewidth]{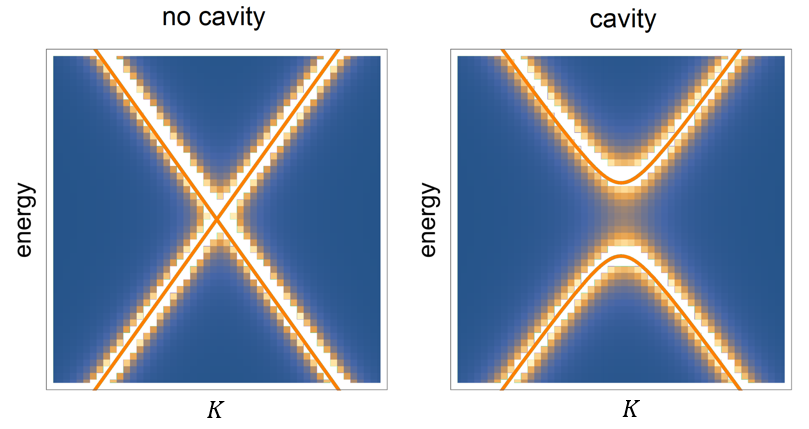}
    \caption{The band structure near the Dirac point for monolayer graphene with time reversal symmetry (\textbf{left panel}) and broken time reversal symmetry (\textbf{right panel}). Units are arbitrary. The orange solid line is the result given by the simple theoretical model based on Eq.\eqref{fig:graphene}. The color map represents the result of coupling monolayer graphene to a chiral optical cavity and computing numerically the spectral function $A(\epsilon,\mathbf{q})$ as outlined in the main text for TBG.}
    \label{fig:graphene}
\end{figure}

For monolayer graphene, near the Dirac point, the effective Hamiltonian can be expressed as,
\begin{equation}
H(\mathbf{q})=v_F(\sigma_x q_x +\sigma_y q_y),
\end{equation}
where $\mathbf{\sigma}$ are the Pauli matrices. The Hamiltonian gives a linear dispersion relation $\epsilon(\mathbf{q})=\pm v_F \sqrt{q_x^2+q_y^2}$ as shown in the left panel of Fig.\ref{fig:graphene}. Time reversal symmetry breaking can be modelled by allowing for a new term in the Hamiltonian whose strength is parameterized by an effective parameter $\tau$,
\begin{equation}
H(\mathbf{q})=v_F(\sigma_x q_x +\sigma_y q_y)+\tau \sigma_z. \label{graphenemodel}
\end{equation}
Solving the eigenvalue problem for the above Hamiltonian, the dispersion relation under broken time reversal symmetry reads $\epsilon(\mathbf{q})=\pm v_F \sqrt{q_x^2+q_y^2+\tau^2}$. The resulting band structure acquire a band gap $\Delta=2v_F \tau$ at the Dirac point, as shown in the right panel of Fig.\ref{fig:graphene}.

In Fig.\ref{fig:graphene} we show the band structure of graphene obtained using the effective model based on Eq.\eqref{fig:graphene} and the one obtained numerically by coupling graphene to a chiral cavity, as explained in the main text for TBG. The two results match very well and show that the chiral cavity opens a gap at the Dirac point as well, as a natural consequence of time reversal symmetry breaking.

\subsection{The role of the cavity frequency}
\begin{figure}[ht]
    \centering
    \includegraphics[width=0.7\linewidth]{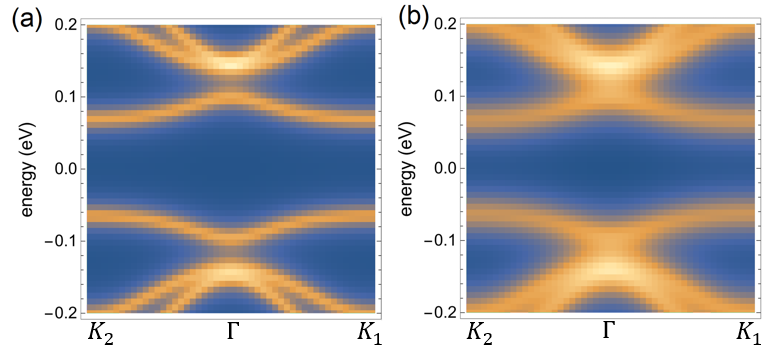}
    \caption{The spectral function $A(\epsilon,\mathbf{q})$ showing the band structure of TBG in an optical chiral cavity for $g=2$ and $\theta=1.5^{\circ}$. Panels \textbf{(a)} and \textbf{(b)} correspond respectively to $\omega_c=0.8$eV, $\omega_c=1.5$eV. These figures must be compared with panel (d) in Fig.\color{red}2 \color{black} which shows the same spectrum for $\omega_c=0.3$eV.}
    \label{fig:3SI}
\end{figure}

In order to complete our analysis, we now discuss briefly the role of the characteristic frequency of the chiral optical cavity $\omega_c$. In the main text, we have fixed $\omega_c=0.3$eV, which is a reasonable and experimentally realizable value. In Fig.\ref{fig:3SI}, we study how the electronic spectrum is modified for larger values of such a frequency. Panels (a) and (b) of Fig.\ref{fig:3SI} have to be compared with panel (d) of Fig.\color{red}2 \color{black} in the main text, which shows the same results for smaller $\omega_c$. We observe that increasing the cavity frequency the electronic bands becomes broader and the quasi-flat bands get closer to the other gapped bands at higher energy. Eventually, as a consequence of this strong broadening, the quasi-flat bands are not isolated anymore. This indicates that a large optical cavity frequency is detrimental in engineering isolated topological flat bands.

\subsection{The efficiency of the cavity}
In order to discuss the effect of the chiral cavity in more detail, we can define the ``flattening-efficiency'' as, 
\begin{equation}
    \eta(g,\theta)=1-\dfrac{\Delta \epsilon(g,\theta)}{\Delta \epsilon(0,\theta)},
\end{equation}
where $\Delta \epsilon(g,\theta)$ is the bandwidth parameter discussed in the main text. From this definition it can be found that $\eta$ takes values in the range from $0$ to $1$. The minimum value $\eta=0$ means that the chiral cavity has no effect on the flatness of the isolated bands. On the contrary, the maximum value $\eta=1$ correspond to a perfect flattening of the band after having introduce the cavity, $\Delta \epsilon(g,\theta)=0$. More in general, a large value of $\eta$ implies that the bands are strongly flattened by the chiral cavity with respect to their structure at $g=0$.

\begin{figure}
    \centering
    \includegraphics[width=0.5\linewidth]{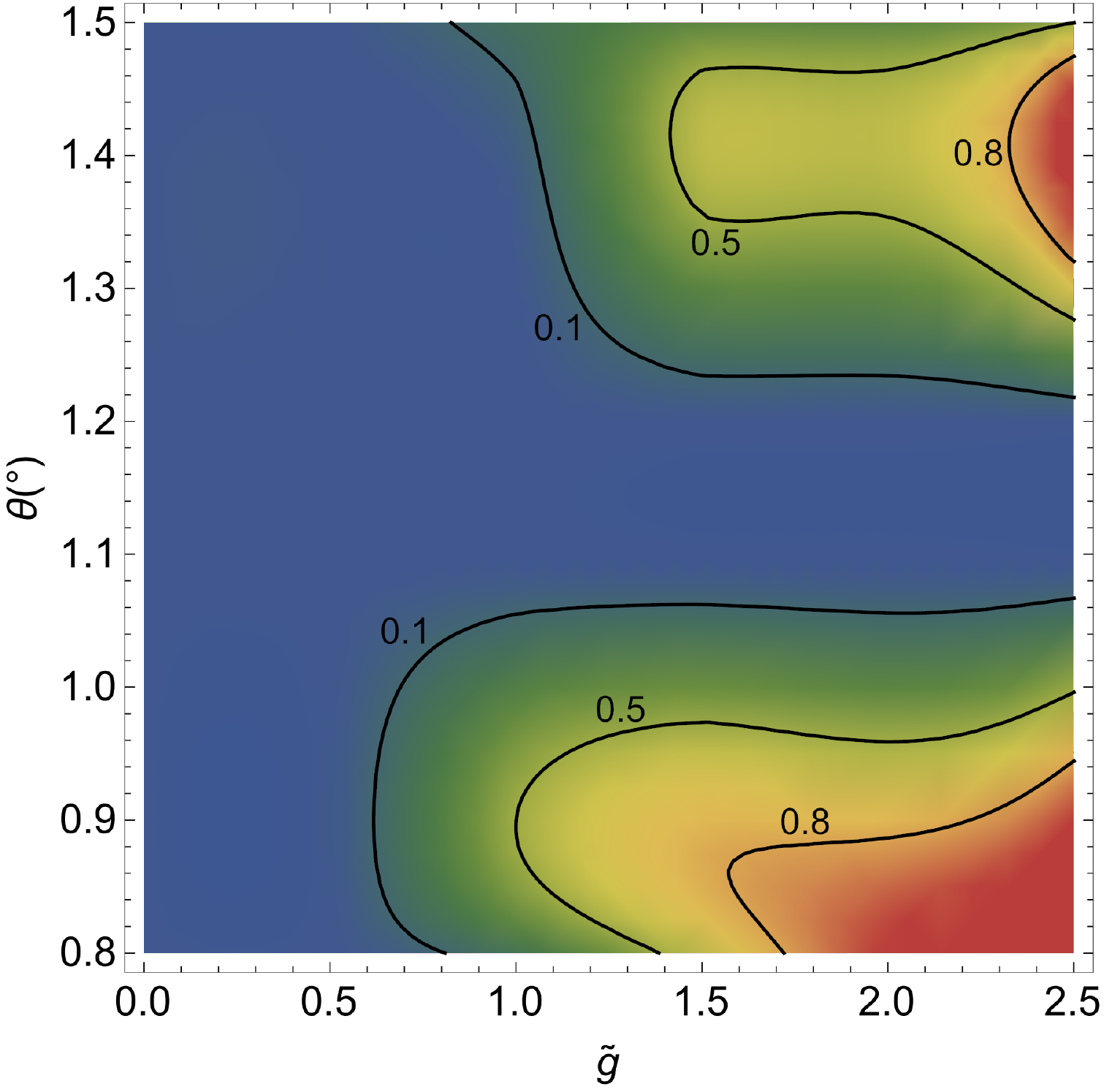}
    \caption{The flattening efficiency parameter $\eta$ as a function of the dimensionless light-matter coupling $\tilde g$ and the twisting angle $\theta$. From blue to red, the efficiency grows larger. Because the band is already flat near the magic angle, the efficiency drops to zero in there.}
    \label{fig:efficiency}
\end{figure}

Fig.\ref{fig:efficiency} shows the flattening parameter $\eta$ as a function of the twisting angle $\theta$ and the dimensionless coupling $\tilde g$ across the phase diagram, around the magic angle value $\theta=1.1^{\circ}$. We observe that, except for small $\tilde g$ or near the magic angle, the chiral cavity can always greatly improve the flatness of band. The results in Fig.\ref{fig:efficiency} are consistent with those reported in the main text in Fig.\color{red}4 \color{black}.

\subsection{Berry curvature and Chern number}
In order to characterize the topological properties of our system, we resort to the analysis of the Berry curvature and the Chern number \cite{RevModPhys.82.1959}.

The Berry curvature describes the geometric curvature in phase space, and is related to the geometric phase acquired by a quantum state as it evolves adiabatically along a closed path in momentum space. In analogy with electromagnetism, the Berry curvature plays the role of a magnetic field in momentum space and it is derived from the curl of the Berry connection, that is the analogous of the gauge field or vector potential. In the two dimensional case, the Berry curvature has only one non-trivial component in the direction perpendicular to 2D plane. 

In wave vector space, the Berry curvature of the $n$th band at wave-vector $\boldsymbol{q}$ is given by,  
\begin{equation}\label{dede}
    \Omega_n (\boldsymbol{q})=i[\langle\partial_{q_x} n(\boldsymbol{q})|\partial_{q_y} n(\boldsymbol{q})\rangle-\langle\partial_{q_y},n(\boldsymbol{q})|\partial_{q_x} n(\boldsymbol{q})\rangle],
\end{equation}
where $|n(\boldsymbol{q})\rangle$ indicates the eigenstate of $n$th band at wave vector $\boldsymbol{q}$. Eq.\eqref{dede} is nothing else than the curl of the Berry connection, $\Omega_n (\boldsymbol{q})=\nabla_{\boldsymbol{q}} \times A_n (\boldsymbol{q})$.

Finally, we can define the topological Chern number $\mathcal{C}$,
\begin{equation}
    \mathcal{C}_n=\frac{1}{2\pi} \int_{\text{BZ}} \Omega_n (\boldsymbol{q}) d\boldsymbol{q},
\end{equation}
where the integration is taken over the whole Brillouin zone (BZ). The Chern number is integer and quantized, $\mathcal{C}=0,\pm 1,\pm 2,\dots$, and play a fundamental role for many transport properties of topological materials.\\

\begin{figure}[htb]
    \centering
    \includegraphics[width=0.4\linewidth]{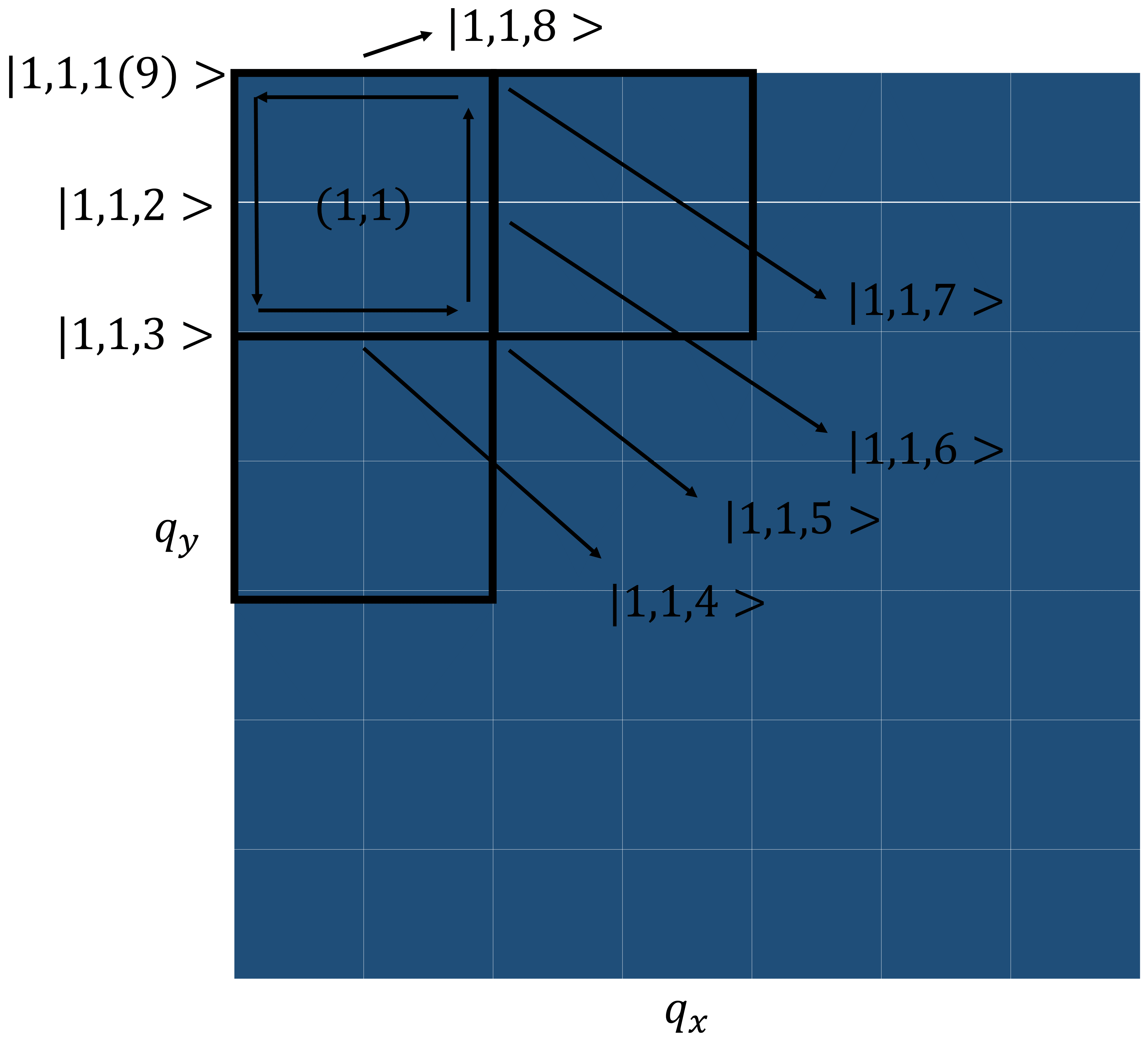}
    \caption{A schematic representation of the algorithm used to compute the Berry curvature and the Chern number. The colored region represents the whole Brillouin zone. The $q$-grid, local loops and label scheme are also indicated.}
    \label{qgrid}
\end{figure}

In our numerical computations, we use the method described in \url{https://topocondmat.org/w4_haldane/ComputingChern.html}. Please see Fig.\ref{qgrid} for computational details.

The local loops can be labeled by a pair of integer indexes $(i,j)$. In each local loop, there are $8$ $q$-grid points on the boundary labeled as $| i,j,l \rangle$ with $l=[1,9]$ oriented clockwise. Because the local loops have only $8$ states, we imposed periodic boundary conditions $| i,j,1 \rangle = | i,j,9 \rangle$. The states involved and labeled here are only those that appear at the boundary of local loops. The average Berry curvature within the local loop $(i,j)$ is given by,
\begin{equation}
  \bar  \Omega_n (i,j) = \mathrm{Arg}\left[\prod_{l=1}^8 \langle i,j,l|i,j,l+1\rangle\right].
\end{equation}
The Chern number of the $n$th band is then given by,
\begin{equation}
    C_n = \sum_{i,j} \Omega_n (i,j).
\end{equation}
This methods was used to compute all the Chern numbers displayed in the main text and figures therein.

\subsection{Large versus small twisting angle}
The band structure of TBG, and consequently the structure of the Berry flux, are very sensitive to the twisting angle. In this section, we will investigate this point further by considering a situation with a large twisting angle and one very close to the magic angle.

\begin{figure}[htb!]
    \centering
    \includegraphics[width=0.8\linewidth]{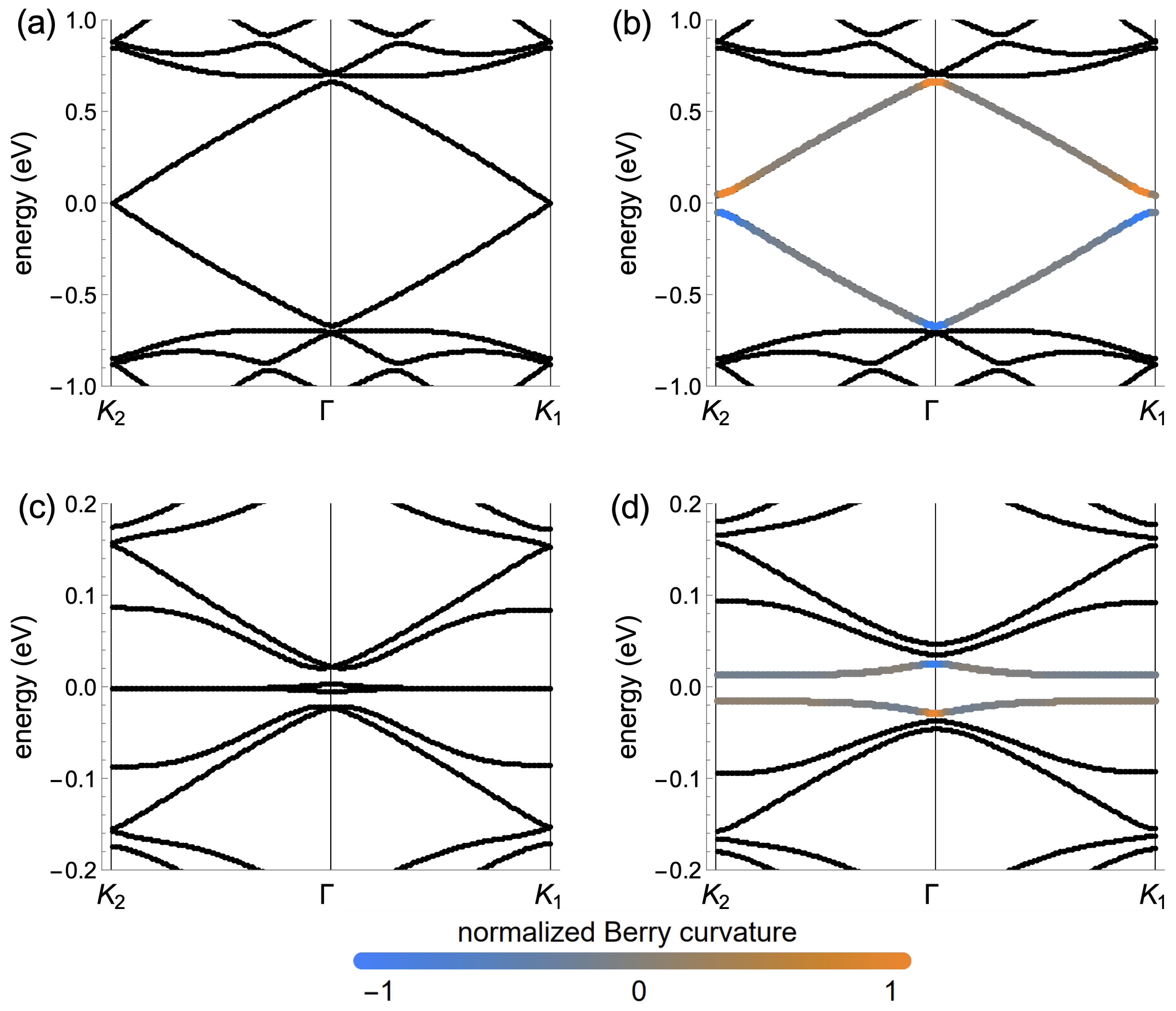}
    \caption{The band structure of TBG at twisting angles $\theta=5^{\circ}$ (top) and $\theta=1.05^{\circ}$ (bottom). The left column shows the results with $\tau=0$ (time reversal symmetry unbroken). The right column corresponds to a soft breaking of time reversal symmetry, $\tau=0.05$, enough to open a small gap. Our band results are consistent with those reported in \cite{doi:10.1073/pnas.1108174108}.}
    \label{q11}
\end{figure}

The band structure in the ``large angle'' regime is reproduced in the top panel of Fig.\ref{q11} for $\theta=5^\circ$ and it displays two clear Dirac cones at both $K_1$ and $K_2$, consistent with previous results \cite{doi:10.1073/pnas.1108174108}. This is similar to what found in \cite{PhysRevResearch.1.023031} for $\theta=7.34^\circ$. In this case, the Berry curvature is peaked at both the $K$ points but at the $\Gamma$ point as well, as checked explicitly in panel (b) of Fig.\ref{q11}. In the presence of time-reversal symmetry, the Berry curvature is an odd function of momentum, \textit{i.e.}, $\Omega(\bold k)=-\Omega(-\bold k)$, implying that at the $\Gamma$ point (\textit{i.e.}, $k=0$), $\Omega(\bold k=0)=0$. Conversely, when a time-reversal symmetry-breaking factor is introduced, no symmetry principle dictates that the Berry curvature must vanish at the $\Gamma$ point.  

The situation exhibits significant differences closer to the magic angle, particularly for the specific case of $\theta=1.5^\circ$ discussed in the main text and reproduced in the bottom panel of Fig. \ref{q11}.
Under these circumstances, and in line with previous computations \cite{doi:10.1073/pnas.1108174108}, the band structure becomes very flat near $K_1$ and $K_2$, and other bands approach the lowest-energy flat bands at the $\Gamma$ point, at which the Berry curvature is peaked. On the contrary, the Berry curvature at the $K$ points becomes very flat. \\

\subsection{The topological phase transition in the TBG-cavity system}
\begin{figure}
    \centering
    \includegraphics[width=0.68\linewidth]{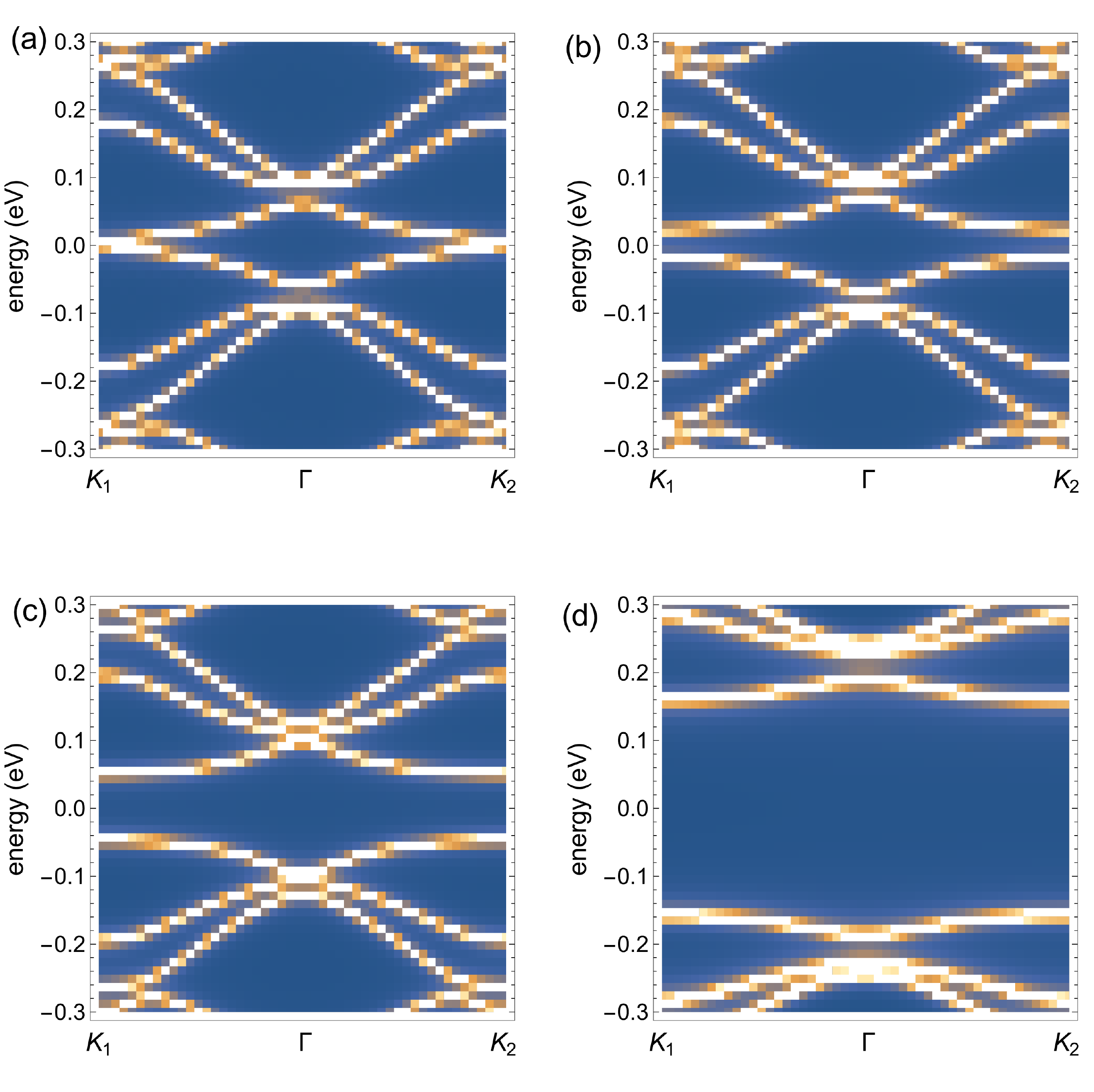}
    \caption{The spectral function $A(\epsilon,q)$ showing the band structure of TBG at the twisting angle $\theta=1.5^\circ$ with light-matter coupling strength $\tilde g=0$, $\tilde g=1$, $\tilde g=1.5$, $\tilde g=2$ for panel (a) to (d) respectively. The optical cavity characteristic frequency is fixed to $\omega_c=0.3$eV.}
    \label{qmiss2}
\end{figure}

In order to make sure that the toy model presented in the main text correctly reflects the behavior of the band structures obtained for the TBG-cavity system, we computed the spectral function $A(\epsilon,q)$ for different light-matter coupling strengths, playing the role of the time reversal symmetry breaking parameter $\tau$. The results are shown in Fig.\ref{qmiss2} and have to be compared with Fig.3 in the main text. The structure of the electronic bands and its behavior upon increasing the strength of time reversal symmetry breaking (TRSB) are compatible and indeed very similar. This analysis further confirms the existence of a topological phase transition induced by TRSB that is accompanied by a gap closing at the $\Gamma$ point as shown in panel (c) of Fig.\ref{qmiss2}. This topological phase transition deserves further study.

\subsection{Experimental feasibility of our proposal}
The proposed cavity size is quite conservative, with an effective mode volume of approximately $(1~\mathrm{\mu m})^3$. Notably, experiments have demonstrated the possibility of building much smaller cavities with an effective mode volume on the order of $V_{\rm eff}\sim 1\times 10^{-5} \times(\lambda/2\sqrt{\epsilon})^{3}$, where $\lambda$ represents the wavelength of the confined photonic mode \cite{scalari2013ultrastrong}. These cavities have been utilized in experiments to investigate Landau polaritons \cite{PhysRevB.90.205309}. Due to their proven experimental feasibility, these effective parameters were also widely adopted in various theoretical papers \cite{schlawin2019cavity,PhysRevB.99.235156}.
 
\begin{figure}
    \centering
    \includegraphics[width=0.36\linewidth]{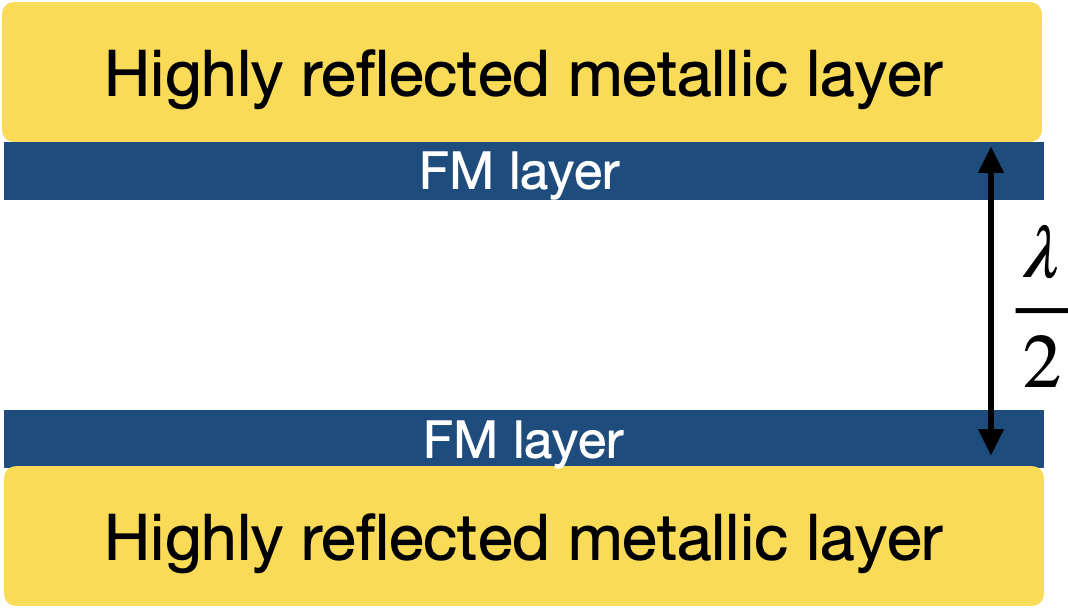}
    \caption{A geometrical demonstration of an experimentally realizable chiral Fabry-P{\'e}rot cavity with small mode volume and time-reversal symmetry breaking.}
    \label{fig:chiralCavity}
\end{figure}

By definition, a chiral cavity breaks time-reversal symmetry, favoring only one handedness of photons. A straightforward method for achieving an effective chiral cavity involves utilizing a Faraday rotator (\textit{e.g.}, ferromagnetic layer) in conjunction with high-quality metallic mirrors to establish the cavity, as illustrated in Figure \ref{fig:chiralCavity}.  
Such chiral cavities have already been realized in experiments \cite{hubener2021engineering}, and a detailed theoretical analysis of these cavities can be found, for example, in \cite{PhysRevA.107.L021501}.

Finally, we remark that our proposed setup does not require a high $Q$ factor, as the mechanism in our proposal does not depend on resonant light-matter coupling. This significant distinction (compared to quantum optics setups) enhances the feasibility of utilizing a cavity to control condensed matter systems. Recently, a plethora of experimental works has emerged to explore cavity many-body systems. Among these experiments, two notable works stand out. One demonstrates that a vacuum cavity can break down the topological protection of quantum hall systems \cite{appugliese2022breakdown}, while the second illustrates that a cavity can be employed to control the metal-to-insulator transition of transition metal dichalcogenide material $\rm Ta S_2$ \cite{jarc2023cavity}.

All in all, the existing evidence supports the experimental feasibility of our proposal.

\end{document}